\documentclass[]{aa} 

\usepackage{graphicx}	
\usepackage{amsmath}	
\usepackage{amssymb}	
\usepackage{comment}
\usepackage{soul}
\usepackage{color}
\usepackage{xcolor}
\usepackage{hyperref}
\usepackage{placeins}
\newcommand{\bl}[1]{\mbox{\boldmath$ #1 $}}

\begin{document}

\title{Dust enrichment and growth in the earliest stages of protoplanetary disk formation}


   \author{ Eduard I. Vorobyov\inst{1}, Vardan G. Elbakyan\inst{2}, Alexandr Skliarevskii\inst{3}, Vitaly Akimkin\inst{4}, and Igor Kulikov\inst{5}}
    \institute{ University of Vienna, Department of Astrophysics, T\"urkenschanzstrasse 17, 1180, Vienna, Austria;    \email{eduard.vorobiev@univie.ac.at}   
    \and 
  Fakultät für Physik, Universität Duisburg-Essen, Lotharstraße 1, D-47057 Duisburg, Germany 
  \and 
       Research Institute of Physics, Southern Federal University, Rostov-on-Don 344090, Russia
  \and
  Institute of Astronomy, Russian Academy of Sciences, Pyatnistkaya str., Moscow, 119017, Russia
  \and
Institute of Computational Mathematics and Mathematical Geophysics SB RAS, Lavrentieva ave., 6, Novosibirsk, 630090 Russia
  }

  \date{}

   \titlerunning{Dust enrichment and growth of protoplanetary disks}
   \authorrunning{Vorobyov et al.}

  \abstract
  {}
 { We numerically investigated dust enrichment and growth during the initial stages of protoplanetary disk formation. A particular objective was to determine the effects of various growth barriers, mimicked by imposing a series of upper permissible limits on maximum dust sizes. } 
  {We used the Formation and Evolution of Stars and Disks on nested meshes (ngFEOSAD) code to simulate the three-dimensional dynamics of gas and dust under the polytropic approximation, from the gravitational collapse of a slowly rotating Bonnor-Ebert sphere to $\approx 12$~kyr after the first hydrostatic core and disk formation.}
  {We found that dust growth begins in the contracting cloud in the evolution stage that precedes disk formation, and that the disk begins to form in an environment already enriched with grown dust. The efficiency of dust growth in the disk is limited by dust growth barriers.  
  For dust grains with maximum sizes $< 100$~$\mu$m, electrostatic or bouncing barriers likely dominate, whereas fragmentation and drift barriers are more important for larger grains.  The disk midplane quickly becomes enriched with dust, while the vertically integrated dust distribution shows notable local variations around the canonical 1:100 dust-to-gas mass ratio. These positive and negative deviations are likely caused by local hydrodynamic flows, as the globally integrated dust-to-gas ratio deviates negligibly from the initial 1:100 value.
  We note that care should be taken when using models with fixed dust sizes, as disks exhibit profound negative radial gradients in dust size even during the earliest stages of disk formation. Models with a constant Stokes number may be preferable in this context.}
{Early dust enrichment and growth may facilitate planet formation, as suggested by observations of protoplanetary disk substructures.}


   \maketitle

\section{Introduction}

Dust enrichment and growth during protoplanetary disk evolution are key steps in planet formation. When these crucial steps towards planet formation occur is not yet fully understood. Dust in the interstellar medium is submicron in size \citep{Mathis1977}, whereas protoplanetary disks in the early T Tauri stage may already host protoplanets, as implied by the presence of deep gaps in gas and dust radial density distributions \citep{2015ALMABrogan,VanDerMarel2019}. This requires dust growth by many orders of magnitude over several hundred thousand years, starting from gravitational contraction of a prestellar cloud and ending with cloud dissipation and accretion onto a young stellar object.   

Growing evidence suggests that the initial stages of dust enrichment above the interstellar value of 1:100 may already occur in the stage that precedes disk formation. This stage is characterized by the gravitational contraction of a prestellar cloud, followed by the formation of the first hydrostatic core (hereafter, FHSC) as defined by Larson \citep{Larson1985}. Observations of young stellar objects suggest the presence of dust grains up to 100~$\mu$m in protostellar envelopes \citep{2019Galametz,Valdivia2019}.  The prevailing explanation for the presence of such large grains involves disk winds, which can lift grains that have already experienced some growth in the disk to hundreds of astronomical units above the disk midplane \citep{2021Tsukamoto,2023Koga,Morbidelli2024}. 
Numerical modeling of dust growth in protostellar envelopes supports this scenario, suggesting that dust may grow to only a few microns \citep{Akimkin2022}. 

However, limitations in current grain growth models suggest that some microphysical processes, not yet fully understood, may still allow dust to grow in situ to grains larger than $1.0$~$\mu$m more rapidly than expected \citep{Morbidelli2024}. For instance, \citep{Bate2022} found that dust growth to $\le 100$~$\mu$m can occur in the inner several hundreds of astronomical units, and in particular in the FHSC. 
The FHSC is not only a transitional structure preceding the protostar -- whose formation begins when the temperature in the interior of the FHSC reaches a threshold value for molecular hydrogen dissociation \citep{2000Masunaga} -- but also a transitional phase toward the protoplanetary disk, which can form around the FHSC even earlier than around the protostar itself \citep{Inutsuka2012,Tomida2015}. Dust grains may dynamically decouple from the gas already within the protostellar envelope, and the resulting differential collapse of dust with respect to the gas can enrich the protoplanetary disk from the moment of its formation \citep{Tsukamoto2017,Cridland2022}.

Rapid dust growth in the earliest stages of disk formation has been reported in several recent studies. For example, \citet{Marchand2023} found that, starting from submicron radii, grain sizes reach more than 100~$\mu$m in an inner protoplanetary disk that is only one thousand years old. Similar results have been reported by \citet{Bate2022} and \citet{VorobyovKulikov2024}. 
Further dust growth and enrichment during the protoplanetary disk stage can be hindered by dust growth barriers caused by electrostatic repulsion, bouncing, fragmentation, and drift \citep{Birnstiel2024}. Incorporating these barriers self-consistently in dust growth models coupled with disk hydrodynamic simulations is a complex task that requires calculating the ionization balance and dust charging, internal grain structure, and dust-gas dynamics. The fragmentation barrier is perhaps the simplest to consider by prescribing a dust fragmentation velocity, which sets the upper dust size in the so-called fragmentation-limited regime. 
However, large uncertainties in fragmentation velocity \citep{Wada2009,2015ApJ...798...34G,Blum2018} lead to significant variations in dust disk properties, depending on the actual value of this parameter \citep{Vorobyov2023a}. 
Dust growth and drift in hydrodynamic models can be further simplified by considering dust dynamics for a range of fixed dust grain sizes (or fixed Stokes numbers) \citep[e.g.,][]{2004RiceLodato,2024MNRAS.528.2490R}, or via the steady-state terminal velocity approximation \citep[e.g.,][]{2020Lebreuilly}. However, it remains unclear when assumptions of fixed dust size or Stokes number are valid. 

Given the complicated nature of dust growth barriers, we approach the problem of the maximum achievable dust size using a simplified framework. We conduct a series of controlled numerical experiments on dust dynamics and growth in the pre- and immediate post-disk formation phases using the three-dimensional Formation and Evolution Of Stars and Disks on nested grids (ngFEOSAD) code.  Dust is allowed to grow through mutual collisions, but its maximum size is artificially capped between 10~$\mu$m and 100~cm. We examine the results to draw conclusions that can be useful for inferring the relative importance of various growth barriers and also assess the applicability of focused simulations of dust dynamics with fixed dust size or Stokes number. We also identify and discuss the initial sources of dust enhancement.

The paper is organized as follows. Section~\ref{Sect:model} describes the hydrodynamic model used to simulate the earliest stages of disk formation. The main results are presented in Sect.~\ref{sect:results}. Model caveats are described in Sect.~\ref{Sect:caveats} and the conclusions are summarized in Sect.~\ref{Sect:conclude}.

\section{Model description}
\label{Sect:model}
The simulations were performed using the ngFEOSAD code. 
The relevant equations and solution methods are thoroughly described in \citep{VorobyovKulikov2024}. 
Here, we briefly review the main concepts and equations; additional details can be found in the aforementioned work.
Using ngFEOSAD, we computed the co-evolution of gas and dust
during the initial stages of protoplanetary disk formation, starting from the gravitational collapse of a slowly rotating prestellar Bonnor-Ebert sphere in full three-dimension. 
We adopted a barotropic equation of state, which included the initial isothermal collapse stage followed by the FHSC and disk formation phases, up to the hydrogen dissociation stage, at which the protostar begins to form.  We terminated our simulations prior to protostar formation because this process could not be fully resolved in our three-dimensional simulations, which had a limited spatial resolution ($\ge 0.14$~au in the central regions). Furthermore, we focused on processes occurring in the forming disk rather than in protostellar interiors, for which lower-dimensional numerical models are better suited \citep{Kuiper2020}. 
We also did not introduce the sink cell, as the results may be sensitive to its specific implementation \citep{2014Machida, 2019VorobyovSkliarevskii, 2020Hennebelle}, which requires careful consideration and is currently under development.

\subsection{Gas component}
The dynamics of gas were followed by solving the equations of continuity and momentum:
\begin{equation}
\label{eq:cont}
\frac{{\partial \rho_{\rm g} }}{{\partial t}}   + \nabla  \cdot 
\left( \rho_{\rm g} {\bl v} \right)  = 0,  
\end{equation}
\begin{equation}
\label{eq:mom}
\frac{\partial}{\partial t} \left( \rho_{\rm g} {\bl v} \right) +  \nabla \cdot \left( \rho_{\rm
g} {\bl v} \otimes {\bl v} + \mathbb{I} P \right)   =    - \rho_{\rm g} \, \nabla \Phi 
 - \rho_{\rm d,gr} \, {\bl f},
\end{equation}
where $\rho_{\rm g}$ is the volume gas density, $\Phi$ is the gravitational potential, $\bl v$ is the gas velocity,  $P$ is the gas pressure, $\mathbb{I}$ is the unit tensor, and $\bl f$ is the friction force per unit dust mass between gas and dust.  

We used the following barotropic equation for the closure relation between gas pressure and density:
\begin{equation}
P_k = c_{\rm s,0}^2 \, \rho_{\rm g}^{\gamma_k} \prod_{i=1}^{k-1}\rho_{{\rm c},i}^{\gamma_i-\gamma_{i+1}},
\,\,\, \mathrm{for} \,\,\, \rho_{{\rm c},k-1} \le \rho_{\rm g} < \rho_{{\rm c}, k} \, 
\label{eosvol},
\end{equation}
where $c_{\rm s,0}=\sqrt{{\cal R} T_{\rm in}/\mu}$ is the initial isothermal sound speed, $T_{\rm in}$ is the initial gas temperature, $\cal R$ is the universal gas constant, and $\mu=2.33$ is the mean molecular weight.  The indices $i$ and $k$ distinguish the four individual components of the piecewise form shown in Table~\ref{table:1}. The first component corresponds to the initial isothermal stage of cloud core collapse, while the remaining three components describe the post-collapse nonisothermal evolution with various ratios of the specific heat $\gamma_{i(k)}$ \citep{2000Masunaga,2015MachidaNakamura}.
We note that for $k=1$, the product term is unity by definition and the pressure reduces to $P_1=c_{\rm s}^2 \rho_{\rm g}^{\gamma_1}$. The gas pressure $P$ in Equation~(\ref{eq:mom}) was set equal to $P_{k}$, where the index $k$ was defined by the condition that $\rho_{\rm g}$ falls between the corresponding critical densities, $\rho_{{\rm c},k-1} \le \rho_{\rm g} < \rho_{{\rm c}, k}$.

\begin{table}
\center
\caption{Parameters of the barotropic relation.}
\label{table:1}
\renewcommand{\arraystretch}{1.5}
\begin{tabular*}{\columnwidth}{ @{\extracolsep{\fill}} c c c }
\hline \hline
$k$ & $\gamma_{i(k)}$ & $n_{{\rm c},{i(k)}}$  \\
\hspace{1cm} & & (cm$^{-3}$)   \\ [0.5ex]
\hline \\ [-2.0ex]
1 & 1.00 & $5{\times}10^{10}$  \\
2 & 1.6667 & $2.5{\times}10^{12}$  \\
3 & 1.4 & $1.5\times 10^{15}$ \\
4 & 1.1 & ... \\ [1.0ex]
\hline
\end{tabular*}
\center{ \textbf{Notes.} The coefficients $\gamma_i$ and $n_{{\rm c},i}$ were taken from \citet{2015MachidaNakamura}. The number density $n_{{\rm c},i}$ relates to the volume density $\rho_{{\rm c},i}$ as $\rho_{{\rm c},i}=\mu m_{\rm H} n_{{\rm c},i}$, where $m_{\rm H}$ is the mass of hydrogen atom.}
\end{table}

\subsection{Dust component}

In this study, dust is treated as a pressureless fluid (see Appendix~\ref{App:one}) consisting of two populations: small dust that is dynamically linked to gas, and grown dust that can dynamically decouple from the gas. The governing equations for their dynamics equations are:

\begin{equation}
\label{eq:cont_dust_small}
\frac{{\partial \rho_{\rm d,sm} }}{{\partial t}}   + \nabla  \cdot 
\left( \rho_{\rm d,sm} {\bl v} \right)  = - S(a_{\rm max}),  
\end{equation}
\begin{equation}
\label{eq:cont_dust_grown}
\frac{{\partial \rho_{\rm d,gr} }}{{\partial t}}   + \nabla  \cdot 
\left( \rho_{\rm d,gr} \, {\bl u} \right)  = S(a_{\rm max}),  
\end{equation}
\begin{eqnarray}
\label{eq:mom-dust}
\frac{\partial}{\partial t} \left( \rho_{\rm d,gr}\, {\bl u} \right) +  \nabla \cdot \left( \rho_{\rm
d,gr} \, {\bl u} \otimes {\bl u}  \right)  &=&     - \rho_{\rm d,gr} \, \nabla \Phi + \rho_{\rm d,gr} \, {\bl f}  \nonumber \\ 
&+&  S(a_{\rm max}) \, {\bl v},
\end{eqnarray}
where $\rho_{\rm d,sm}$ and $\rho_{\rm d,gr}$ represent the volume densities of small and grown dust, respectively, $\bl u$ is the velocity of grown dust, and $S(a_{\rm max})$ is the conversion rate between small and grown dust populations.

This two-component dust model was originally developed by \citet{2018VorobyovAkimkin} with modifications to include the gas back-reaction introduced by \citet{2018Stoyanovskaya}.
Small dust is defined as grains with sizes between $a_{\rm min}=5\times 10^{-3} \ \mu \rm m$  and $a_{*} = 1 \ \mu \rm m$, while grown dust comprises aggregates ranging in size from $a_{*}$ to $a_{\rm max}$. The value of $a_\ast$ separating small dust and grown dust is chosen to ensure that grown dust dynamically decouples from, and small dust dynamically couples to, gas in the protoplanetary disk environment \citet[][see Appendix D]{Vorobyov2022}. However, in the collapsing cloud core, dust-to-gas decoupling may occur at smaller than $a_\ast=1.0$~$\mu$m dust sizes due to lower gas densities. The dependence of our results on the choice of $a_\ast$ will be considered in a follow-up paper. Another planned improvement is to incorporate small dust dynamics using the terminal velocity approximation \citep[e.g.,][]{2014LaibePrice,2017LinYoudin}.

Initially, the total dust mass in a collapsing prestellar cloud was partitioned as 70\%:30\% between the small and grown dust populations, respectively (see Sect.~\ref{Sect:init}). As the collapse proceeds and the disk forms and evolves, small dust can grow and transition into the grown dust population. 
In both components, dust is assumed to follow a size distribution given by a simple power law: 
\begin{equation}
N(a) = C \cdot a^{ - {\rm p}},
\label{eq:dustdistlaw}
\end{equation} where $C$ is a normalization constant and ${\rm p} = 3.5$.  In this study, the power index $\rm p$ is kept constant, although it can vary, provided the form of this variation is known from other more sophisticated dust growth models.

\subsection{Dust growth model} 
The maximum value of the dust size $a_{\rm max}$ is calculated using a simple monodisperse dust growth model and can vary in space and time. 
The following equation describes the time evolution of $a_{\rm max}$: 
\begin{equation}
{\partial a_{\rm max} \over \partial t} + ({\bl u} \cdot \nabla ) a_{\rm max} = \cal{D},
\label{eq:dustA}
\end{equation}
where the rate of dust growth due to collisions and coagulation is computed using a monodisperse model \citep{2012Birnstiel}:
\begin{equation}
\cal{D} = {\rho_{\rm d} \mathit{u}_{\rm rel} \over \rho_{\rm s}}.
\label{eq:dustrate}
\end{equation}
This rate includes the total volume density of dust $\rho_{\rm d}=\rho_{\rm d.sm}+\rho_{\rm d,gr}$, the dust material  density $\rho_{\rm s}$, and the relative velocity of grain-to-grain collisions defined as $\mathit{u}_{\rm rel} = (\mathit{u}_{\rm br}^2 + \mathit{u}_{\rm turb}^2)^{1/2}$, where $\mathit{u}_{\rm br}$ and $\mathit{u}_{\rm turb}$ account for the Brownian and turbulence-induced local motion of dust grains, respectively.  In particular, $u_{\rm turb}$ makes the largest contribution to dust growth in our model and is expressed as \citep{2007OrmelCuzzi}:
\begin{equation}
        u_{\rm{turb}} = \sqrt{{3 \alpha \over \mathrm{St}+\mathrm{St}^{-1}}} c_{\rm s},
    \label{turb_vel}
\end{equation}
where $c_{\rm s}$ is the adiabatic speed of sound, $\alpha$ is the parameter describing the strength of turbulence in the disk, and $\mathrm{St}$ is the Stokes number defined as $\mathrm{St}=t_{\rm stop} \Omega_{\rm K}$. Here, $\Omega_{\rm K}$ is the Keplerian angular velocity and $t_{\rm stop}$ is the stopping time. 
The $\alpha$ parameter describes the level of turbulence in the disk and is set to $10^{-3}$, implying low turbulence, which is consistent with observations \citep{2023Rosotti}. 
Determining $\Omega_K$ on the fly during simulations proved to be nontrivial. We attempted to use the local angular velocity as a proxy of $\Omega_{\rm K}$ but this approach performed poorly in disks perturbed by spiral density waves and mass infall from the envelope.
As in our previous work \citep{VorobyovKulikov2024},  $\Omega_K$  was set equal to that of a 0.05~$M_\odot$ central object, which is roughly the upper mass of the FHSC \citep{Tomida2010}. Considering that our runs were limited to the very early stages of star and disk formation, this should have introduced only small errors to $u_{\rm{turb}}$.

As the maximum size of dust grains $a_{\rm max}$ begins to increase according to the adopted dust growth model, part of the small dust population is converted to grown dust. This process is described by the conversion rate $S(a_{\rm max})$, which can be expressed as
\begin{equation}
    S(a_{\rm max}) = {1 \over \Delta t} \frac
    { \rho_{\rm d,sm}^n \int_{a_*}^{a_{\mathrm{max}}^{\rm n+1}} a^{3-\mathrm{p}}da -
    \rho_{\rm d,gr}^n \int_{a_{\rm min}}^{a_*} a^{3-\mathrm{p}}da 
    }
    {
    \int_{a_{\rm min}}^{a_{\mathrm{max}}^{n+1}} a^{3-\mathrm{p}}da
    },
    \label{growth:rate}
\end{equation}
and is derived assuming the conservation of dust mass and continuous dust size distribution across $a_{*}$ (see Eq.~\ref{eq:dustdistlaw}). Here, $\Delta t$ is the hydrodynamic time step and the indices $n$ and $n+1$ denote the current and next hydrodynamic steps of integration, respectively. The detailed derivation can be found in \citet{Molyarova2021} and \citet{Vorobyov2022}.

\subsection{Dust growth barriers}

Dust growth in a protoplanetary disk is hindered by the so-called barriers, including the bouncing barrier, fragmentation barrier, radial drift barrier, and electrostatic barrier \citep{2009Okuzumi,2010ZsomOrmel,Wada2011,Okuzumi2011_charge,Windmark2012,2016Birnstiel, Akimkin2020_charge, 2023ApJ...953...72A}. Each barrier depends sensitively on the local physical conditions in the disk and the dust properties. For instance, the fragmentation barrier increases with the fragmentation velocity squared. This quantity is currently poorly constrained and may vary from 0.5~m~s$^{-1}$ to tens of meters per second \citep{2008ARA&A..46...21B,Wada2009,2015ApJ...798...34G}.  The electrostatic barrier depends on the relative number of dust grains compared to charge carriers (electrons and ions), as well as the temperature and the degree of turbulence in the disk \citep{Akimkin2015}. The bouncing barrier depends on the relative velocity of the colliding dust particles and their structure, particularly the filling factor \citep{Wada2011}. 

Among these barriers, only the radial drift barrier can be accounted for in our simulations self-consistently, as we solve the full equations of gas and dust dynamics, including their mutual friction (see below). However, considering the other barriers self-consistently is problematic. Therefore,  we adopted a simplified approach and imposed a maximum
dust size $a_{\rm max}^{\rm cap}$, which limits dust growth through mutual collisions. 
More specifically, when $a_{\rm max}$, defined by Equation~(\ref{eq:dustA}), exceeds $a_{\rm max}^{\rm cap}$, the growth rate $\mathcal{D}$ from Equation~(\ref{eq:dustrate}) is set to zero and $a_{\rm max}$ is set equal to $a_{\rm max}^{\rm cap}$.

For $a_{\rm max}^{\rm cap}$, we considered six different values: 10~$\mu$m, 100~$\mu$m, 1.0~mm, 1.0~cm, 10~cm, and 100~cm, which are constants of time and space. The lowest value of $a_{\rm max}^{\rm cap}=10$~$\mu$m can be considered as the threshold imposed by the electrostatic barrier, which typically suspends dust growth at several $\mu$m until the barrier is broken, first in the outer cold disk regions \citep{2009Okuzumi, 2010ZsomOrmel, 2011OkuzumiTanaka, Akimkin2015, 2020Steinpilz, 2023Akimkin}. The highest value of $a_{\rm max}^{\rm cap}=100$~ cm may represent the fragmentation barrier, corresponding to the highest fragmentation velocity of several tens of meters per second for compact ice dust aggregates \citep{Wada2009}. The intermediate values of $a_{\rm max}^{\rm cap}$ represent a combination of fragmentation and bouncing barriers for a spectrum of fragmentation velocities and bouncing efficiencies.

\subsection{Drag force}

The drag force between dust and gas in our model is calculated as
\begin{equation}
    {\bl f} = \frac{{\bl v} - {\bl u}}{t_{\rm stop}},
    \label{eq:fric}
\end{equation}
where the stopping time, $t_{\rm stop}$, is given by
\begin{equation}
\label{tstop}
    t_{\rm stop} = \frac{a \, \rho_{\rm s}} {\rho_{\rm g} v_{\rm th}}.
\end{equation}
Here,  $v_{\rm th}$ is the mean thermal velocity of the gas, and $\rho_{\rm s}= 3.0$~g~cm$^{-3}$ is the material density of dust grains. Since grown dust in our model has a spectrum of sizes from $a_\ast$ to $a_{\rm max}$, the size of dust grains $a$ in Equation~(\ref{tstop}) should be weighted over this spectrum. However,  the range between $a_\ast$ and  $a_{\rm max}$ may span several orders of magnitude during disk evolution, and the average dust grain size, given by $\sqrt{a_\ast a_{\rm max}}$,  becomes much smaller than $a_{\rm max}$. 
In this work, we are interested in the dynamics of dust grains that serve as the main dust mass carriers.
For the chosen slope $p=3.5$, large grains near $a_{\rm max}$ predominantly determine the dust mass \citep{2016Birnstiel}.  Therefore, we use the maximum size of dust grains, $a_{\rm max}$, when calculating the values of $t_{\rm stop}$.  We also note that Equation~\ref{tstop} assumes that the dust is in the Epstein drag regime, where the mean free path of hydrogen molecules is much larger than the size of dust grains.  This condition breaks, for example, in the Stokes regime, and the stopping time must be modified accordingly \citep[see, e.g.][]{Vorobyov2023a,Birnstiel2024}. Figure ~\ref{fig:1d_rad_all_mod_2} demonstrates that the Stokes regime occurs only for models with $a_{\rm max}^{\rm cap}\ge 1.0$~cm, and only within the inner several astronomical units occupied by the FHSC. Since our study focuses primarily on dust dynamics and growth in the disk and collapsing envelope, the Epstein drag regime is justified.

\subsection{Distinguishing the disk from the infalling envelope}
\label{Sect:disk-track}
In our numerical simulations, we distinguished the disk from the infalling envelope for the sake of subsequent analysis. To this end, we adopted the disk tracking conditions outlined in \citet{2012Joos}. Specifically, we used the following criteria to determine whether a particular computational cell belongs to the disk:
 \begin{itemize}
     \item the gas rotational velocity must be faster than twice the radial infall velocity, $v_{\phi}>2v_{\rm r}$,
     \item the gas rotational velocity must be faster than twice the vertical infall velocity, $v_{\phi}>2v_{\rm z}$,
     \item gas must not be thermally supported, $\frac{\rho_{\rm g} v_{\phi}^2}{2}>2P$,
     \item the gas number density must be higher than $10^{10}$~cm$^{-3}$.
 \end{itemize}
Here, $v_{\rm z}$, $v_{\rm r}$, and $v_\phi$ are the components of gas velocity in the cylindrical coordinates, $\rho_{\rm g}$ is the volume density of gas, and $P$ is the gas pressure. 
If any of the first three conditions are not met, a particular grid cell is not qualified as belonging to the disk. The fourth condition must always be satisfied. We note that we have increased the threshold for the gas number density to $10^{10}$~cm$^{-3}$, compared to $10^{9}$ in \citet{2012Joos}, as the higher value better describes the disk in our models.

\subsection{Initial and boundary conditions}

\label{Sect:init}
For the initial cloud core configuration, we constructed a Bonnor-Ebert sphere with the central density $3\times 10^4$~cm$^{-3}$.  The ratio of the central density to that at the outer edge of the sphere, a quantity that determines the shape of the sphere and its stability to gravitational collapse, was set to 15, making the sphere marginally unstable. The initial temperature was set equal to 10~K. The ratio of rotational to gravitational energy was $\beta=0.5$\% and the ratio of thermal to gravitational energy is 70\%. 

The initial dust size distribution spans from $5\times10^{-3}$~$\mu$m to 2.0~$\mu$m, as suggested by the MRN (Mathis, Rumpl, Nordsiek) distribution in interstellar clouds \citep{Mathis1977}. The division between small and grown dust grains at $a_\ast=1.0$~$\mu$m is, however, arbitrary. 
This value is motivated by long drift timescales of dust with size $\le 1.0$~$\mu$m in the midplane of a protoplanetary disk \citep{Vorobyov2022}, allowing us to neglect the dynamics of small dust. However, this may not be the case in a collapsing cloud core, as shown in Sect.~\ref{sect:results}, and the effects of smaller $a_\ast$ will be considered in a follow-up study. For the adopted slope of the dust size distribution $p=3.5$, about 70\% of the total dust mass budget is in the form of small dust grains, while the rest is in the form of grown dust. The initial total dust-to-gas mass ratio is 0.01. 

The cloud is initially embedded in an external medium whose pressure is chosen to maintain the Bonnor-Ebert sphere in marginal equilibrium.
To initiate the collapse, we introduced a 30\% positive perturbation to the gas and dust densities. 
The resulting mass of the unstable cloud core is 
$M_{\rm cl} = 0.89~M_\odot$,  and its radius is $R_{\rm cl}=3.9\times10^{-2}$~pc.

\begin{figure*}
    \centering
    \includegraphics[width=2\columnwidth]{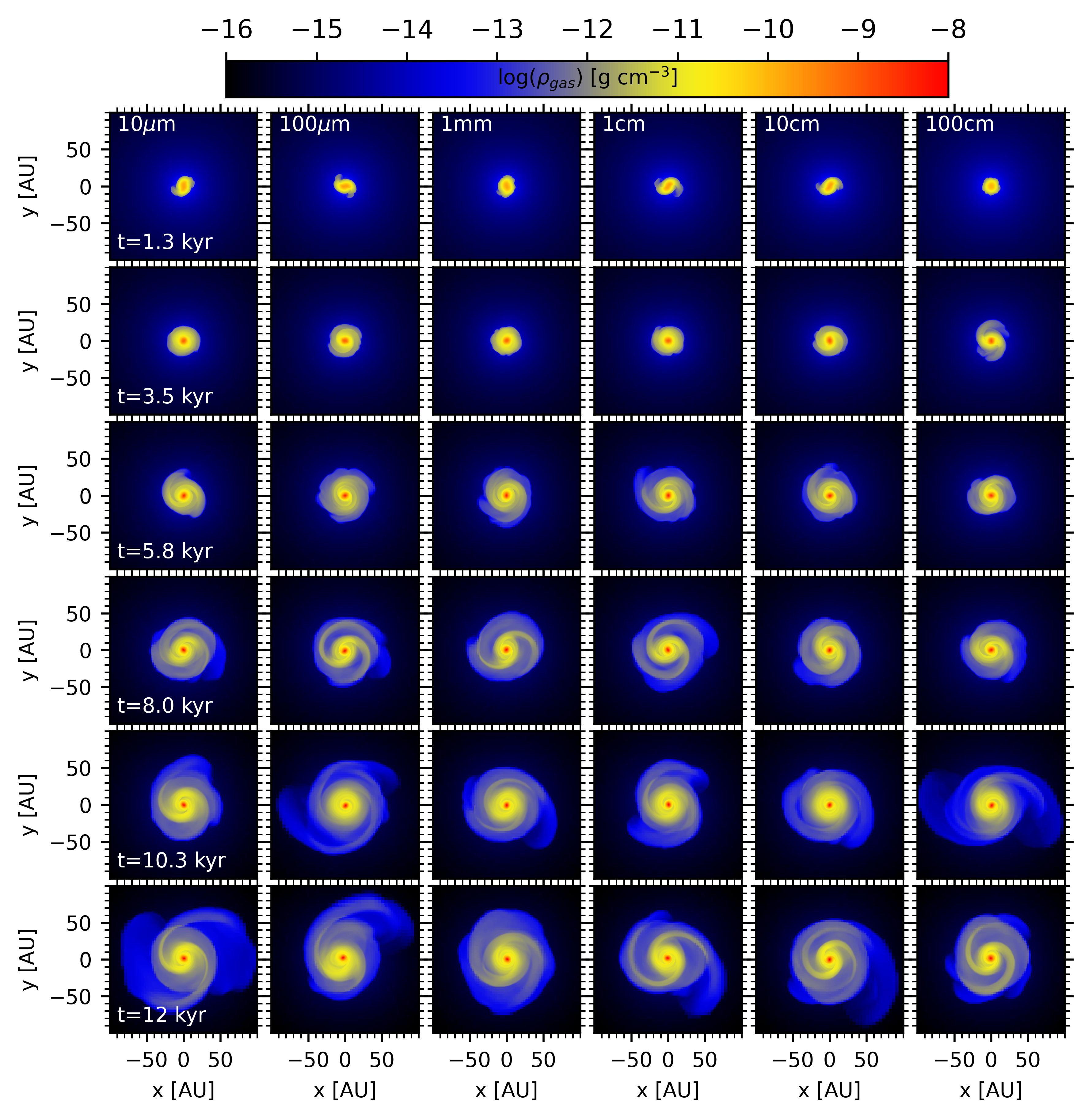}
    \caption{Time evolution of the gas volume density in the disk midplane for all six models considered. Columns from left to right correspond to models with maximum allowed dust sizes of 10~$\mu$m, 100~$\mu$m, 1.0~mm, 1.0~cm, 10~cm, and 100~cm, respectively. The inner spatial region shown measures $200 \times 200$~au$^2$ in size. Time is measured from the disk formation epoch.}
    \label{fig:6x6_gas}
\end{figure*}

 \begin{figure}
    \centering
    \includegraphics[width=1\columnwidth]{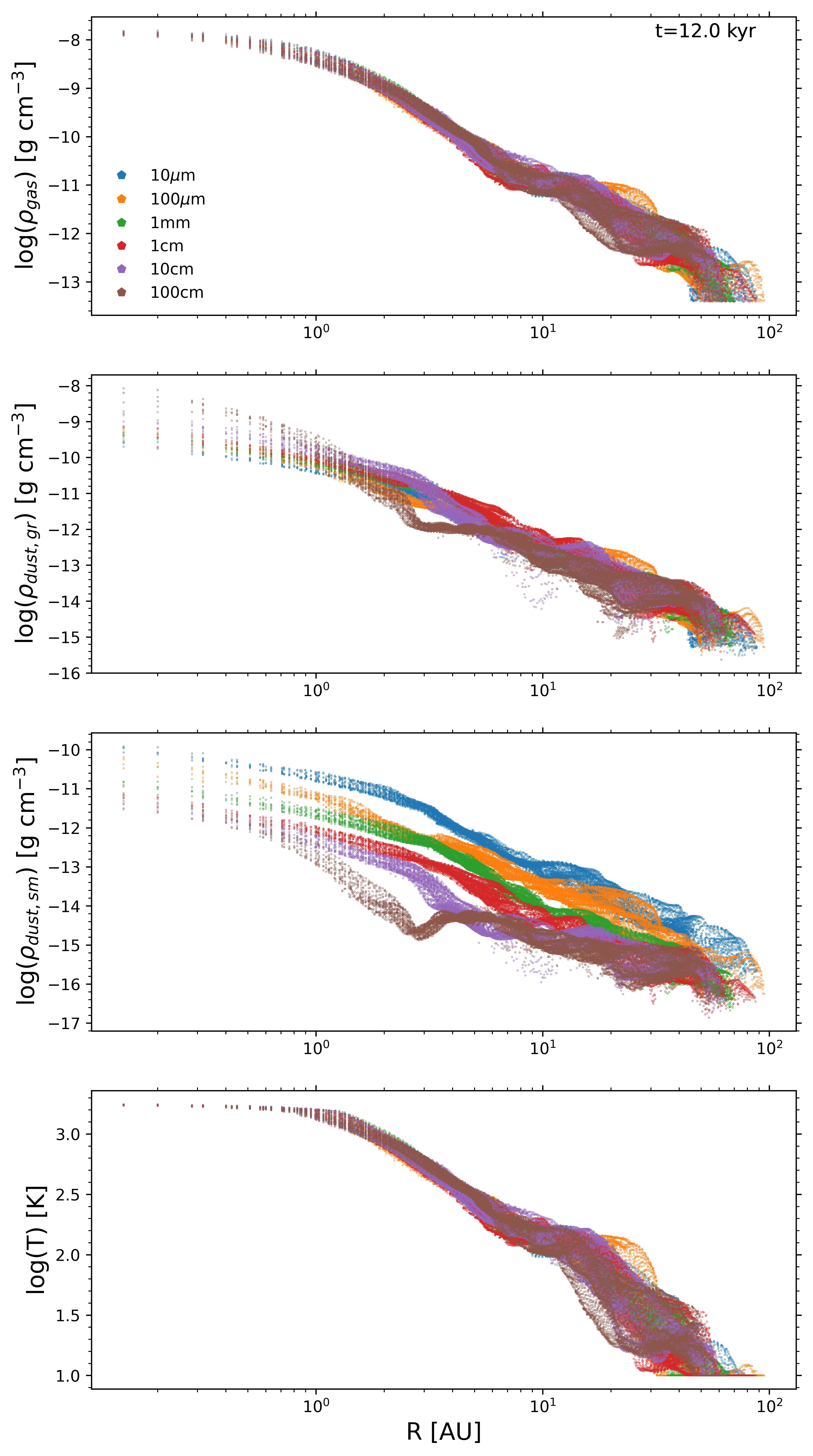}
    \caption{Radial dependence of disk midplane characteristics for all six models. Shown are (from top to bottom) are the gas volume density, grown dust volume density, small dust volume density, and temperature. }
    \label{fig:1d_rad_all_mod}
\end{figure}

\subsection{Numerical solution method}
To solve the hydrodynamic equations for the gas component, we use the Godunov method with the Harten–Lax–van Leer contact (HLLC ) Riemann solver \citep{Toro2019}. We solve the equations for the grown dust component using the Godunov method, similar to that used to solve the hydrodynamic equations but with a one-wave HLL Riemann solver, due to the pressureless nature of the dust fluid.
We treat gravity and friction forces separately after the Godunov step. The gravitational potential is found using the fast Fourier transform and the convolution theorem as described for the case of regular grids in \citet{1987Binney}, with modifications for nested meshes as described in \citet{VorobyovMcKevitt2023}. The method is globally second-order accurate, except at grid interfaces, where we apply first-order interpolation of the boundary values.
We compute the update of gas and dust velocities due to the friction force $\bl f$ between the two components using an analytical solution, under the assumption of constant $t_{\rm stop}$ and a constant dust-to-gas ratio in a given cell during a hydrodynamic time step \citep{2015LorenBate, 2018Stoyanovskaya}.
We calculate the time step using the Courant-Friedrichs-Lewy condition with the Courant number set to 0.25.
A detailed description of the solution procedure and relevant tests are provided in \citet{VorobyovKulikov2024}.

We employed Coarray Fortran for distributed-memory parallelism, which offers a simpler coding model than the conventional message-passing interface (MPI), while being only slightly slower in performance \citep{mcKevitt2024}. The ngFEOSAD code is fast and environmentally friendly. For this work, we used 12 nested levels with 64 cells in each coordinate direction, resulting in approximately 3.15 million grid cells in total and a minimum spatial resolution of $\approx 0.1$~au in the innermost several astronomical units. A typical run-time is 20 days on AMD EPYC 7713 processors with four nodes, totaling 192 cores. This corresponds to approximately 92,000 core hours.

\section{Main results}
\label{sect:results}

 \begin{figure*}
    \centering
    \includegraphics[width=1\columnwidth]{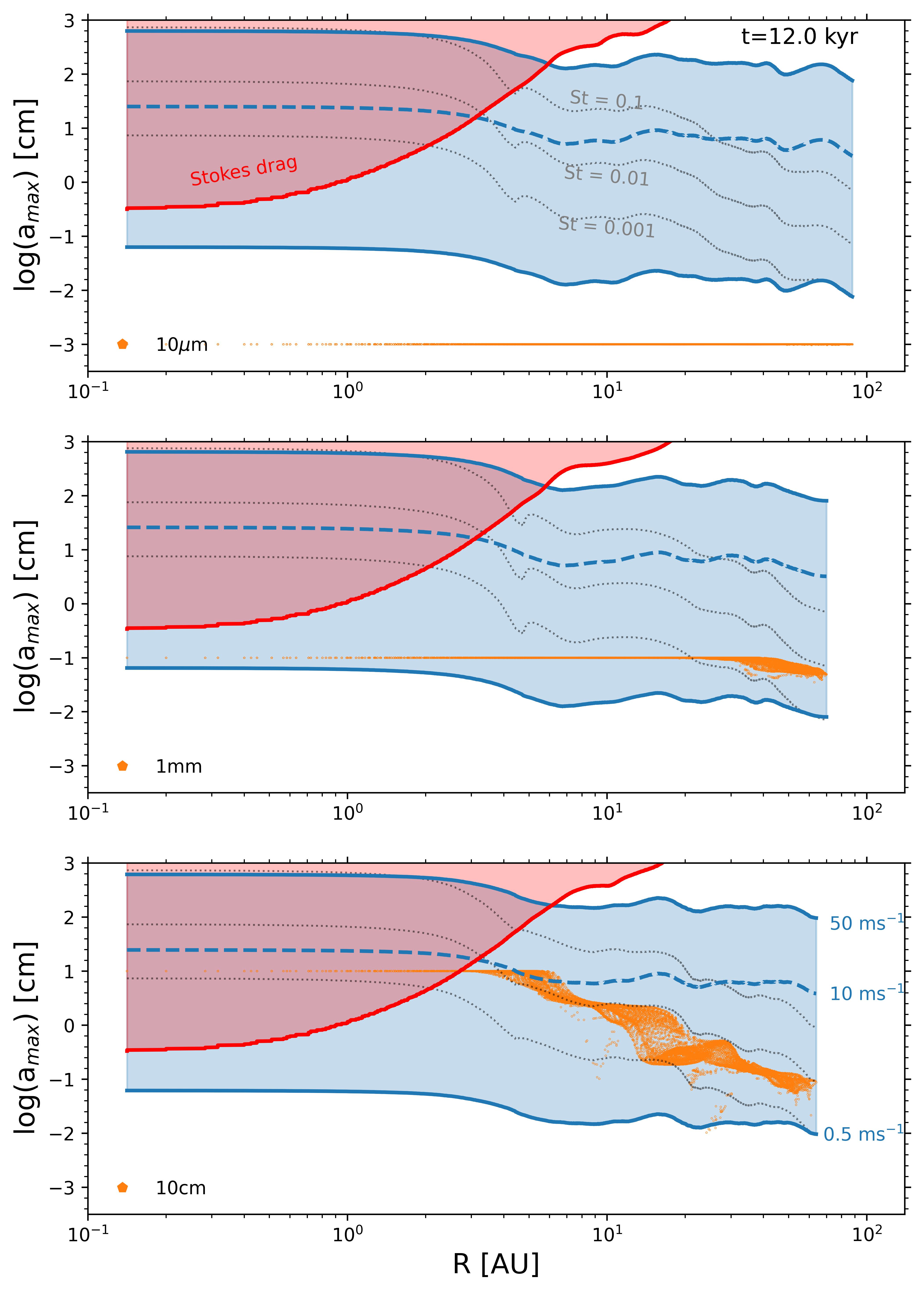}
    \includegraphics[width=1\columnwidth]{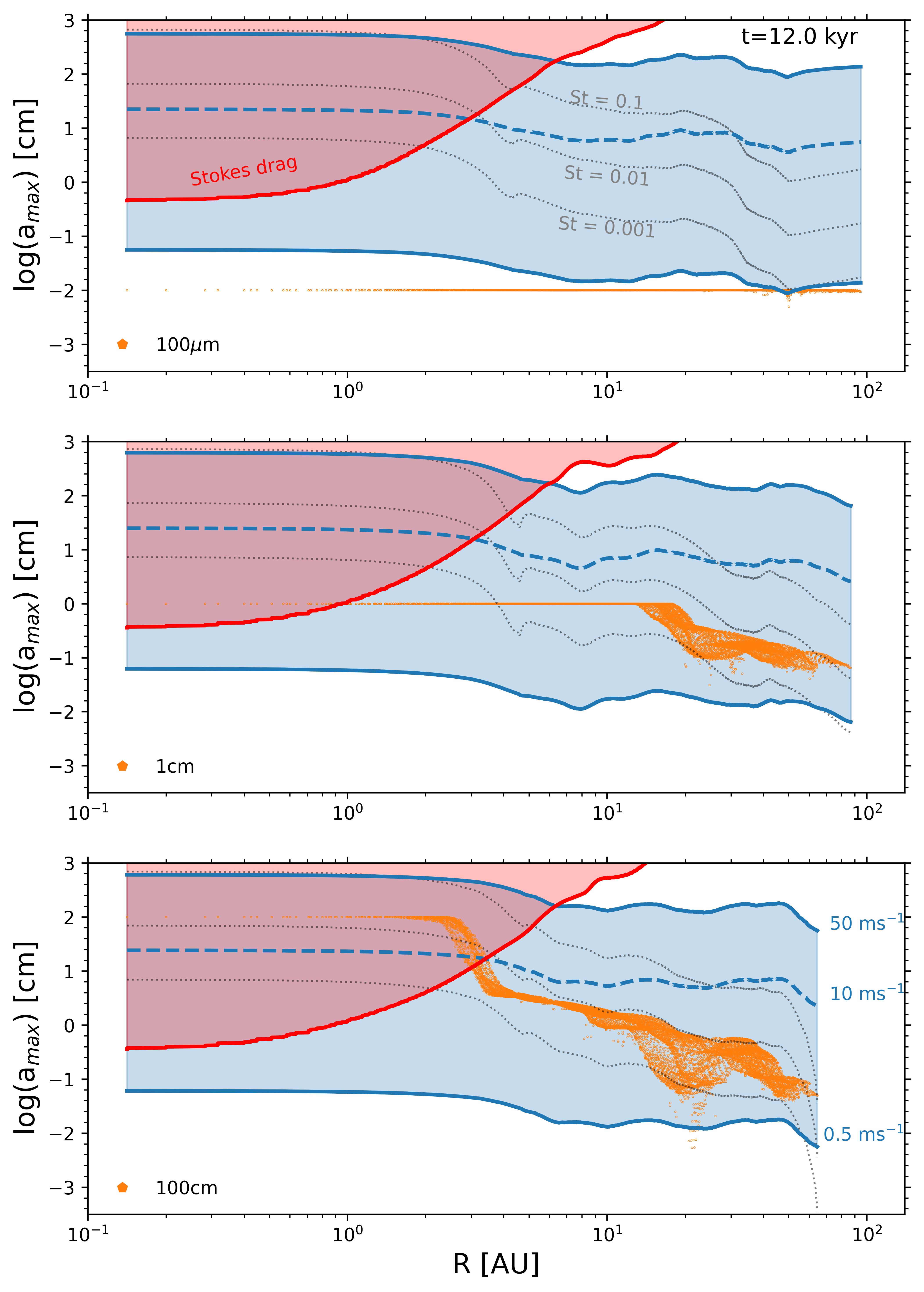}    
    \caption{Radial profiles of  the maximum dust size $a_{\rm max}$ (orange symbols), for all six models with different maximum allowed size, $a_{\rm max}^{\rm cap}$. The six panels (from left to right, top to bottom) correspond to models with $a_{\rm max}^{\rm cap}$=10~$\mu$m, $a_{\rm max}^{\rm cap}$=100~$\mu$m, $a_{\rm max}^{\rm cap}$=1.0~mm, $a_{\rm max}^{\rm cap}$=1.0~cm,  $a_{\rm max}^{\rm cap}$=10~cm, and $a_{\rm max}^{\rm cap}$=100~cm.  Blue curves indicate the maximum dust sizes in the fragmentation-limited dust growth regime for fragmentation velocities of 0.5~m~s$^{-1}$ (lower solid curve), 10~m~s$^{-1}$ (middle dashed curve), and 50~m~s$^{-1}$ (upper solid curve). Black dotted curves illustrate dust sizes for fixed Stokes numbers of 0.1, 0.01, and 0.001 (from top to bottom). The red curve marks the boundary between the Stokes and Epstein drag regimes.}
    \label{fig:1d_rad_all_mod_2}
\end{figure*}

Once we apply a positive density perturbation to the initial Bonnor-Ebert sphere, it contracts under the force of its own gravity and gradually spins up.
The FHSC begins to form near the coordinate center at $\approx 193$~kyr after the start of the simulations, as indicated by positive deviations of the gas temperature from the initial isothermal value of 10~K in this region. By the end of the simulations (204.5~kyr), the temperature at the center of the FHSC reaches $\approx1800$~K, just below the threshold for molecular hydrogen dissociation, the process that leads to the collapse of the FHSC and the formation of the protostar \citep{Masunaga2000}. Our limited numerical resolution does not allow us to follow the formation of the protostar. The transition from FHSC to protostar may occur somewhat earlier than predicted in our model, as we do not fully resolve the inner regions, and the gas temperature is smoothed out over the inner 0.1~au. Radiation transfer simulations are required to better capture the evolution of gas temperature in the FHSC \citep{Tomida2010,Kuiper2018,2023A&A...680A..23A}, a module that is currently under development for ngFEOSAD.

\subsection{General disk evolution analysis}

Figure~\ref{fig:6x6_gas} shows the time evolution of the gas density $\rho_{\rm g}$ in the midplane ($z=0$) for the six models with different maximum dust sizes allowed, $a_{\rm max}^{\rm cap}$.  The disk grows around the FHSC as material with progressively higher angular momentum accretes from the infalling envelope. In these images, the disk is distinguished from the infalling envelope not only by its higher density but also by regions characterized by a profound nonaxisymmetric pattern with a dominant two-armed spiral. These "prestellar" disks are gravitationally unstable, as also seen in other three-dimensional hydrodynamic simulations \citep[e.g.][]{Inutsuka2012,Bate2022}. 
The gravitational instability in such a young disk is maintained by infall from the envelope. This process was first reported in two-dimensional thin-disk simulations by \citet{VorobyovBasu2005}, was further examined in semi-analytic models of \citet{2008Kratter}, and was also detected in HH~111~VLA source \citep{Turner2020}.
Because gravitational instability is a stochastic process, small variations in gas dynamics during the disk formation stage, caused by friction of gas with dust of different sizes, lead to different spiral patterns. However, the overall gas disk evolution remains similar across all considered models.

Figure~\ref{fig:1d_rad_all_mod} shows the radial profiles of various properties of the disk in the midplane ($z=0$) at $t=12$~kyr after disk formation. The data identified as part of the disk, based on the adopted disk tracking mechanism (see Sect.~\ref{Sect:disk-track}), are plotted; the infalling envelope is excluded from the analysis. The radial profiles of gas volume density, grown dust density, and temperature show small variations among models but have similar overall shapes. The formation of the FHSC is most evident in the radial temperature profiles, which show a plateau with a nearly constant temperature of 1800~K extending to 1.0~au.  A similar, though less pronounced, plateau appears in the gas density distribution. The plateau in the grown dust density is least pronounced, especially for large $a_{\rm max}^{\rm cap}$, likely due to inward radial drift.  The small dust density is consistently lower across the entire disk for models with higher $a_{\rm max}^{\rm cap}$, reflecting efficient conversion of small dust into grown dust during the growth process. We conclude that the fragmentation barrier may affect dust in the early stage of disk evolution. 
Gas dynamics may also be indirectly influenced by variations in the ionization fraction, which depends on the surface area of dust grains, and by changes in dust opacity, which is sensitive to the minimum and maximum dust sizes.

\subsection{Radial distributions of maximum dust size and Stokes number}

The orange symbols in Fig.~\ref{fig:1d_rad_all_mod_2} show the maximum dust size in the disk midplane for the six models with different $a_{\rm max}^{\rm cap}$.  We compare our model data with expectations from fragmentation-limited dust growth models. The thick blue curves indicate the maximum
size that dust grains would attain in our models if their collisional growth was limited by fragmentation, assuming fragmentation velocities of $u_{\rm frag}=0.5$~m~s$^{-1}$, 10~m~s$^{-1}$, and 50~m~s$^{-1}$ \citep{Wada2009,Okuzumi2016}. These curves represent the fragmentation barriers for the selected fragmentation velocities. The barriers are calculated similarly to the definition given in \citet{2016Birnstiel}, but using volume densities rather than column densities to account for the three-dimension nature of our modeling: 
\begin{equation}
\label{afrag}
a_{\rm frag} = \frac{\rho_{\rm g}u^2_{\rm frag}}{3 \alpha \rho_{\rm s}  v_{\rm th} \Omega_{\rm K}},
\end{equation}
where $\Omega_{\rm K}$ is the Keplerian angular velocity. 
The thin dotted curves show the dust size that the particles would reach in our models if they were characterized by fixed Stokes numbers: $\mathrm{St}=10^{-3}$, 10$^{-2}$, and 10$^{-1}$. The blue and dotted curves are obtained by calculating the corresponding values across the entire disk and then averaging them along the azimuth at each radial distance. Small-scale variations result from nonaxisymmetries in our model disk, an effect that is absent in the axisymmetric disk model of \citet[][their Fig. 4]{2023ASPC..534..717D}.

\begin{figure}
    \centering
    \includegraphics[width=1\columnwidth]{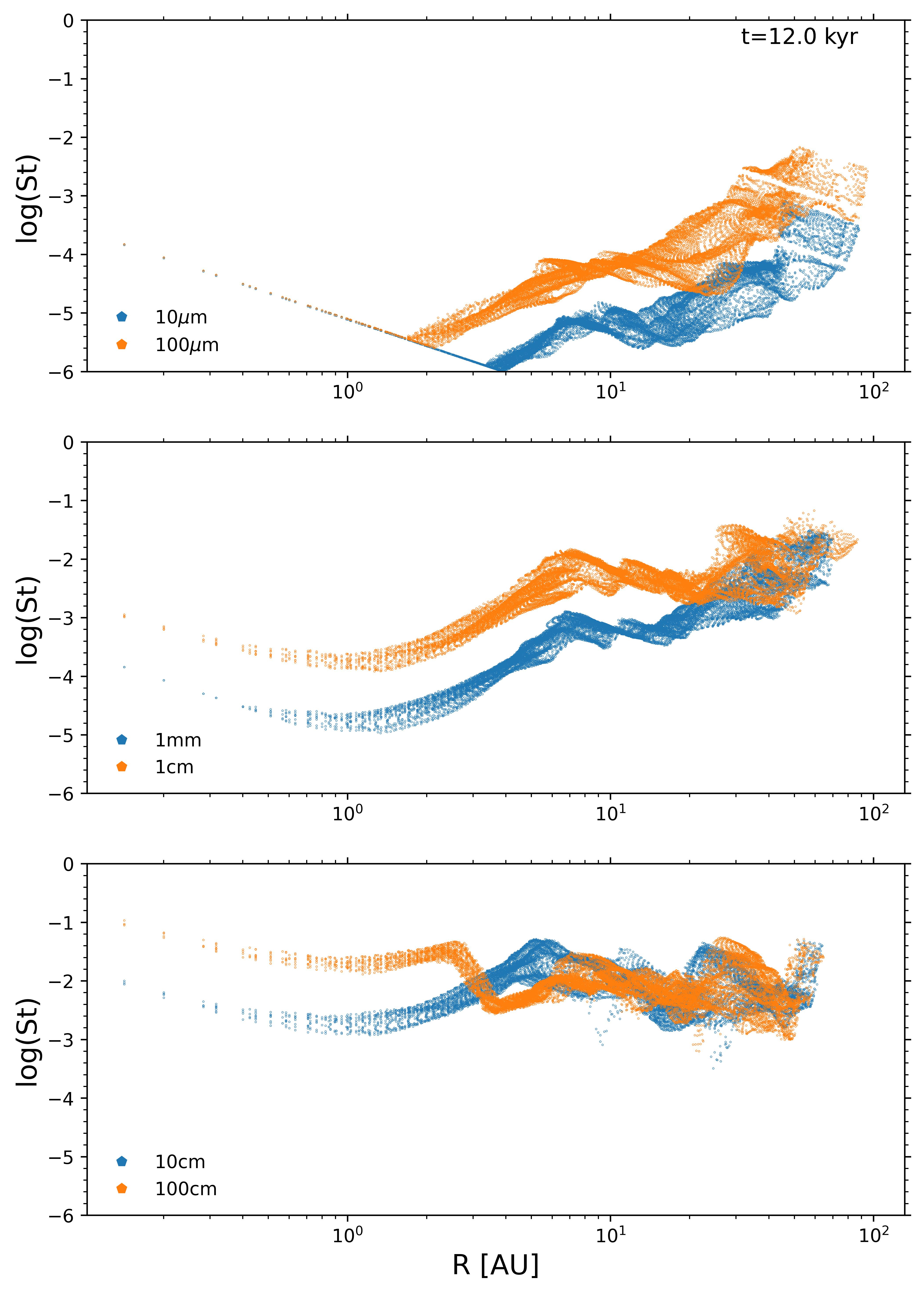}
    \caption{Similar to Fig.~\ref{fig:1d_rad_all_mod} but for the Stokes number that corresponds to $a_{\rm max}$.}
    \label{fig:1d_rad_all_mod_3}
\end{figure}

Several conclusions can be drawn from Figure ~\ref{fig:1d_rad_all_mod_2}, which compares the radial distributions of $a_{\rm max}$ in our models with the fragmentation barriers $a_{\rm frag}$ and the constant $\mathrm{St}$ lines. For $a_{\rm max}^{\rm cap}$=10~$\mu$m, the fragmentation barrier is not reached, even at the lowest considered fragmentation velocity of $u_{\rm frag}=0.5$~m~s$^{-1}$. 
If such small dust particles are inferred from observations of young disks, their growth is likely limited by barriers other than collisional fragmentation,  or the fragmentation velocity is even lower than 0.5~m~s$^{-1}$. These grains have Stokes numbers well below $10^{-3}$, implying that radial drift is insignificant in this case (see also the discussion in Figure~\ref{fig:1d_rad_all_mod_3}).

For $a_{\rm max}^{\rm cap}$=100~$\mu$m,  the fragmentation barrier corresponding to the lowest considered fragmentation velocity of $u_{\rm frag}=0.5$~m~s$^{-1}$  is marginally reached in the disk region from 5-6~au out to the disk's outer edge. 
The presence of dust particles of 100~$\mu$m may therefore imply a low fragmentation velocity near 0.5~m~s$^{-1}$.  In the outermost disk regions, dust particles approach $\mathrm{St}=10^{-3}$, although the bulk of the disk midplane remains below this threshold.
In both cases, $a_{\rm max}^{\rm cap}$=10~$\mu$m and $a_{\rm max}^{\rm cap}$=100~$\mu$m, dust growth is saturated at our imposed upper limit, and the drift barrier is ineffective in setting the maximum dust size.

For $a_{\rm max}^{\rm cap}$=1.0~mm and 1.0~cm,  the maximum dust sizes fall within those defined by the fragmentation barriers at $u_{\rm frag}=0.5$~m~s$^{-1}$ and  $u_{\rm frag}=10$~m~s$^{-1}$. 
In both models, a region appears where dust growth is not saturated ($a_{\rm max}$ is smaller than the corresponding $a_{\rm max}^{\rm cap}$), which implies that radial drift has become efficient and dust grains cannot reach the imposed upper limits on the maximum dust size. Indeed, the Stokes numbers are now above $10^{-3}$, reinforcing our conclusion that radial dust drift is no more negligible. Interestingly, the $a_{\rm max}^{\rm cap}$=1.0~cm model can  be described by a slowly varying $\mathrm{St}$, ranging from $10^{-3}$ to $10^{-2}$, in the disk regions from 10~au outward.

For $a_{\rm max}^{\rm cap}$=10 and 100~cm,  dust growth saturates only in the inner several astronomical units, while the drift-dominated region extends inward and almost reaches the FHSC for the latter model. 
The maximum dust sizes are mainly below those imposed by the fragmentation barrier at $u_{\rm frag}=10$~m~s$^{-1}$, with an exception of the innermost regions, where an even higher $u_{\rm frag}$ is required to account for the values of $a_{\rm max}$. The trend of $a_{\rm max}$ aligning with the lines of constant Stokes number (close to $10^{-2}$) becomes more pronounced for these two models.

If the electrostatic barrier is crossed (perhaps already in the collapse stage), dust can grow to sub-mm, mm, or even cm-sized grains in the disk, prior to protostar formation during the collapse of the FHSC, depending on the details of dust fragmentation and drift processes.  Such rapid dust growth has also been noted in previous studies of disk formation stages \citep[e.g.,][]{2018VorobyovAkimkin,Bate2022}. Furthermore, 
the existence of large grains seems to be confirmed in observations of the inner disk regions in systems with luminosity outbursts \citep{Liu2021,Houge2024}, which can assist planet formation in terrestrial zones. This finding is also relevant for the pebble accretion scenario in planetary core growth \citep{2017JohansenLambrechts}. 

Figure~\ref{fig:1d_rad_all_mod_2} also shows that
$a_{\rm max}$ is characterized by a clear negative radial gradient when $a_{\rm max}^{\rm cap}$ is sufficiently large. Radial dust size gradients have been inferred in protoplanetary disks at later stages of disk evolution by analyzing the radial distribution of dust spectral indices 
\citep[e.g.,][]{Pinte2016,2016A&A...588A..53T}. Our modeling shows that radial differentiation in the maximum dust size can exist even in pre-stellar disks, provided that fragmentation and bouncing are inefficient at limiting dust growth.

Furthermore, disk models studying the dynamics of dust populations with a fixed dust size may only be justified if the dust growth barriers effectively limit the maximum dust size below 100~$\mu$m. 
Otherwise, assuming a constant dust size would not accurately reflect the true radial distribution of dust size in the disk. In other words, dust grains of 1.0~mm or 1.0~cm may not exist throughout the entire disk, and assuming a population of dust grains with a fixed size at these large values may not be justified for the full extent of the disk.

The radial distribution of the Stokes number $\mathrm{St(a_{\rm max})}$, corresponding to the maximum dust size $a_{\rm max}$, shows a different trend. We calculate $\mathrm{St(a_{\rm max})}$ in the disk midplane and show it in Fig.~\ref{fig:1d_rad_all_mod_3} for all six models with different $a_{\rm max}^{\rm cap}$. A positive radial gradient in $\mathrm{St(a_{\rm max})}$ is evident for $a_{\rm max}^{\rm cap}\le 1.0$~mm across most of the disk extent, except in the FHSC, where the trend reverses. For $a_{\rm max}^{\rm cap}\ge 10$~cm, the radial distribution of $\mathrm{St(a_{\rm max})}$ does not exhibit a clear radial trend across the disk and is instead characterized by an order of magnitude scatter at a given radial distance,  as was already observed when analyzing Fig.~\ref{fig:1d_rad_all_mod_2}.  As discussed previously, these models experience minimal restriction on dust growth from fragmentation and/or bouncing barriers, so the use of fixed Stokes number models may be justified in this case. Numerical models that compute the dynamics of multiple dust bins of fixed size or Stokes number without dust growth may benefit from insights gained in our study \citep[e.g.][]{2004RiceLodato,2020Lebreuilly}.
The case in which the entire disk volume is considered, rather than just the disk midplane as discussed above, is further discussed in Appendix~\ref{App:two}.  We note that the negative gradient of $\mathrm{St}$ at $R<4$~au in the $a_{\rm max}^{\rm cap}\le 100$~$\mu$m  models results from setting a lower limit on the stopping time in our simulations and thus is an artificial feature.

Another notable conclusion from Fig.~\ref{fig:1d_rad_all_mod_3} is that the Stokes number rarely exceeds 0.1 in the early disk evolution, even in the model with $a_{\rm max}^{\rm cap}= 100$~cm. As described in our previous work \citep{VorobyovKulikov2024}, this can be partly attributed to the high gas densities and temperatures in young disks, both of which reduce the stopping time of dust grains.  Radial dust drift can also contribute to limiting the Stokes number, especially in models with large $a_{\rm max}^{\rm cap}\ge 10$~cm.  As dust particles grow in size and their Stokes number increases, they begin drifting
toward pressure maxima. The resulting increase in gas temperature and density offsets the increase due to  $a_{\rm max}$ in the definition of the Stokes number.  
This self-regulating effect may also explain why strong radial gradients of the Stokes number diminish in models with increasing $a_{\rm max}^{\rm cap}$. 
The absence of $\mathrm{St}\gg0.1$ in young disks,  at least in the disk midplane, should be considered when studying dust enrichment in spiral arms, which are regions of local pressure maxima \citep{2004RiceLodato,Boss2020,2024MNRAS.528.2490R}. Low $\mathrm{St}$ may hinder this process until a much later stage of disk evolution, when disk density and temperature decrease and $\mathrm{St}$ may begin to approach unity. 

Finally, we note a steep change in $a_{\rm max}$ and $\mathrm{St}$ in the $a_{\rm max}^{\rm cap}\ge 100$~cm at around 3~au. This feature is likely due to efficient radial dust drift towards the pressure-supported  FHSC, which results in a local minimum in the dust-to-gas mass ratio at this radial distance (see Fig.~\ref{fig:xi-total}). This also makes dust growth in these regions strongly drift-dominated.

\begin{figure}
    \centering
   \includegraphics[width=1\columnwidth]{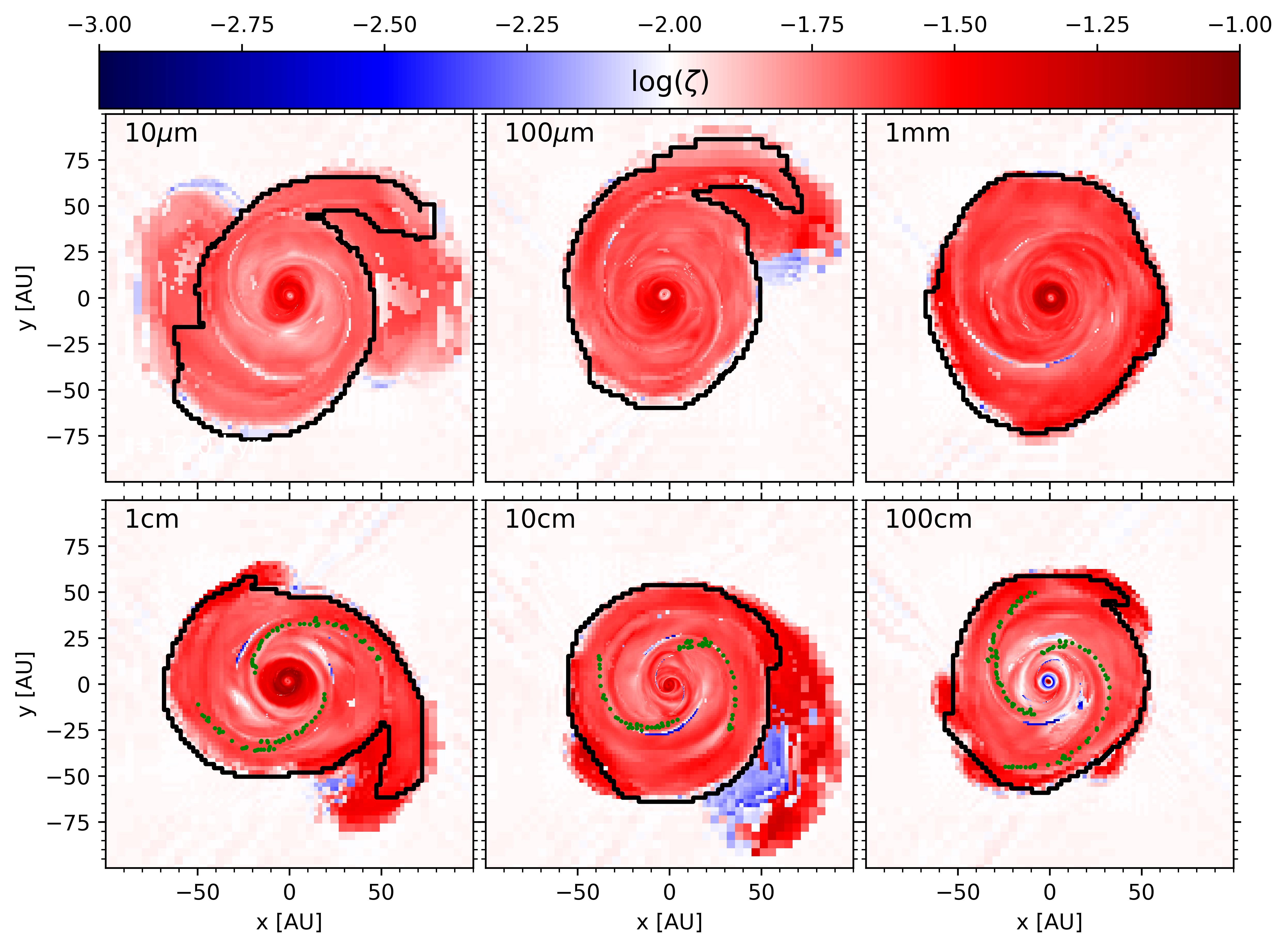}
    \caption{Ratio of total dust to gas volume densities in the disk midplane, $\zeta$, at $t=12$~kyr after disk formation. Six models are shown, each with a different maximum allowed dust size $a_{\rm max}^{\rm cap}$: 10~$\mu$m, 100~$\mu$m, 1.0~mm, 1.0~cm, 10~cm, and 100~cm (arranged from left to right and top to bottom). Black contour curves delineate the disk extent, and green circles indicate the position of the spiral arms in the last three models.}
    \label{fig:2d_dens_dust_tot}
\end{figure}

\begin{figure}
    \centering
    \includegraphics[width=1\columnwidth]{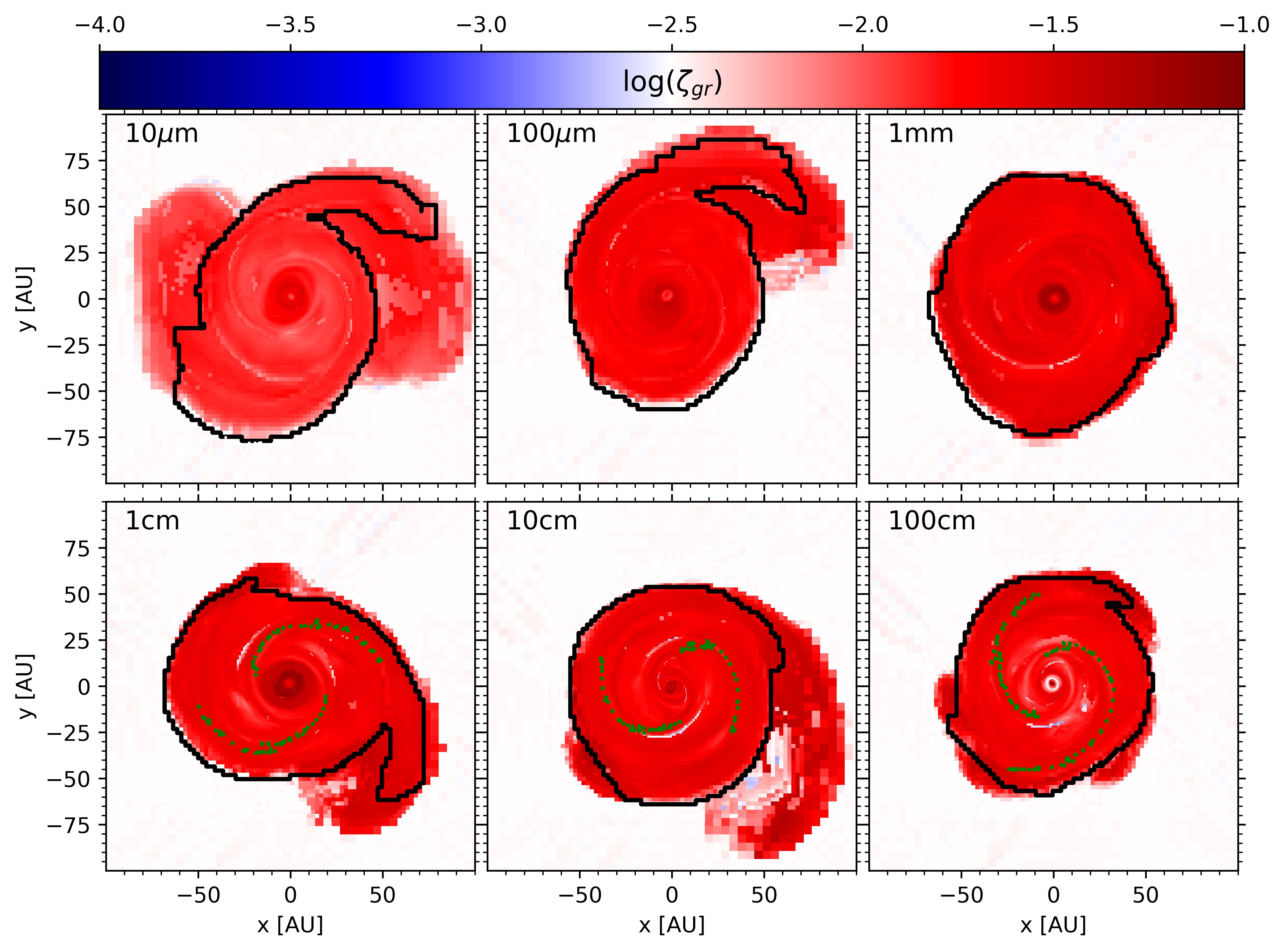}
    \caption{ Ratio of the grown dust to gas volume densities, $\zeta_{\rm gr}$, similar to Fig.~\ref{fig:2d_dens_dust_tot} .}
    \label{fig:2d_dens_dust_grown}
\end{figure}

\subsection{Dust enhancement}

\begin{figure*}
    \centering
    \includegraphics[width=1\columnwidth]{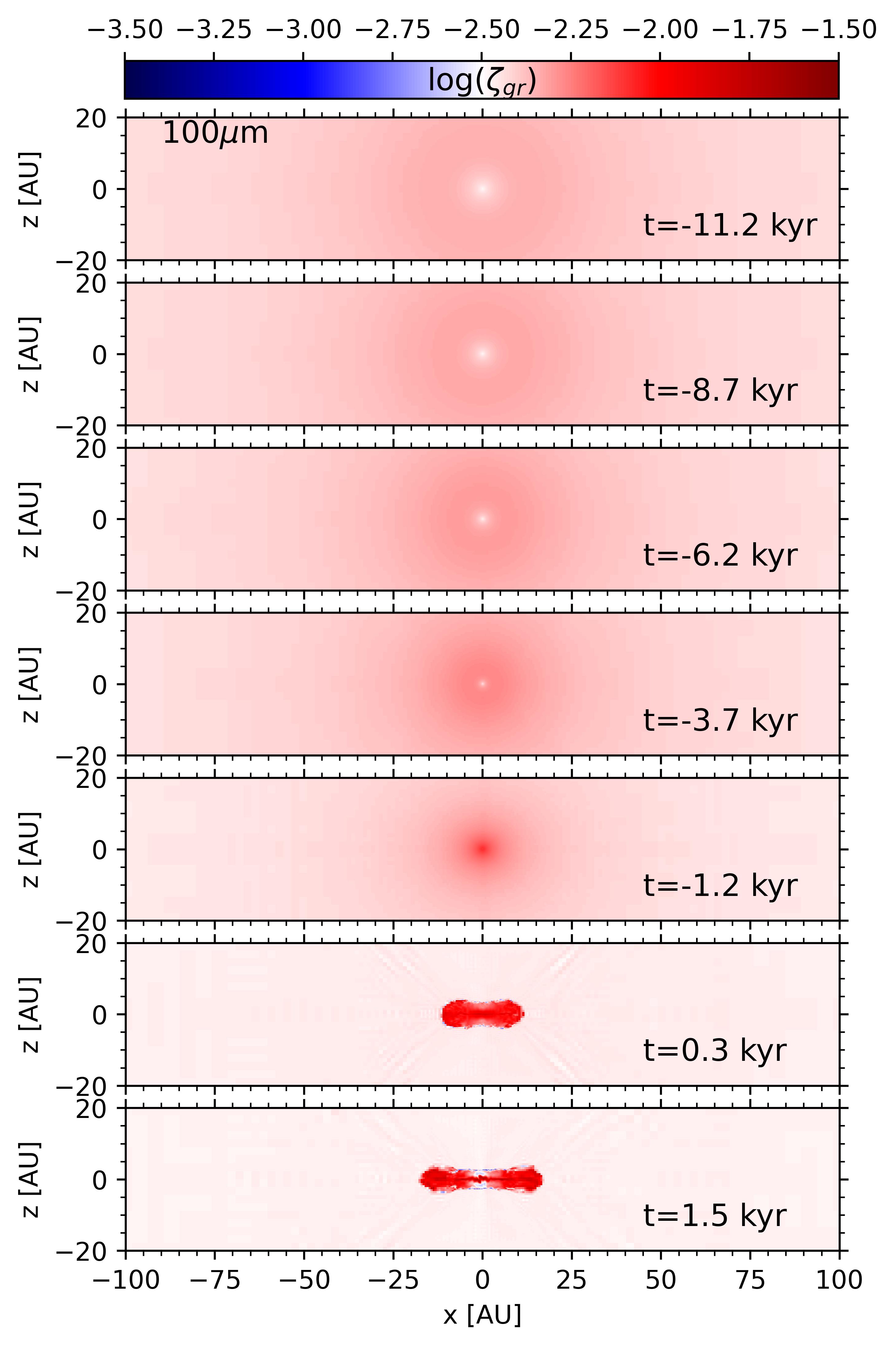}
    \includegraphics[width=1\columnwidth]{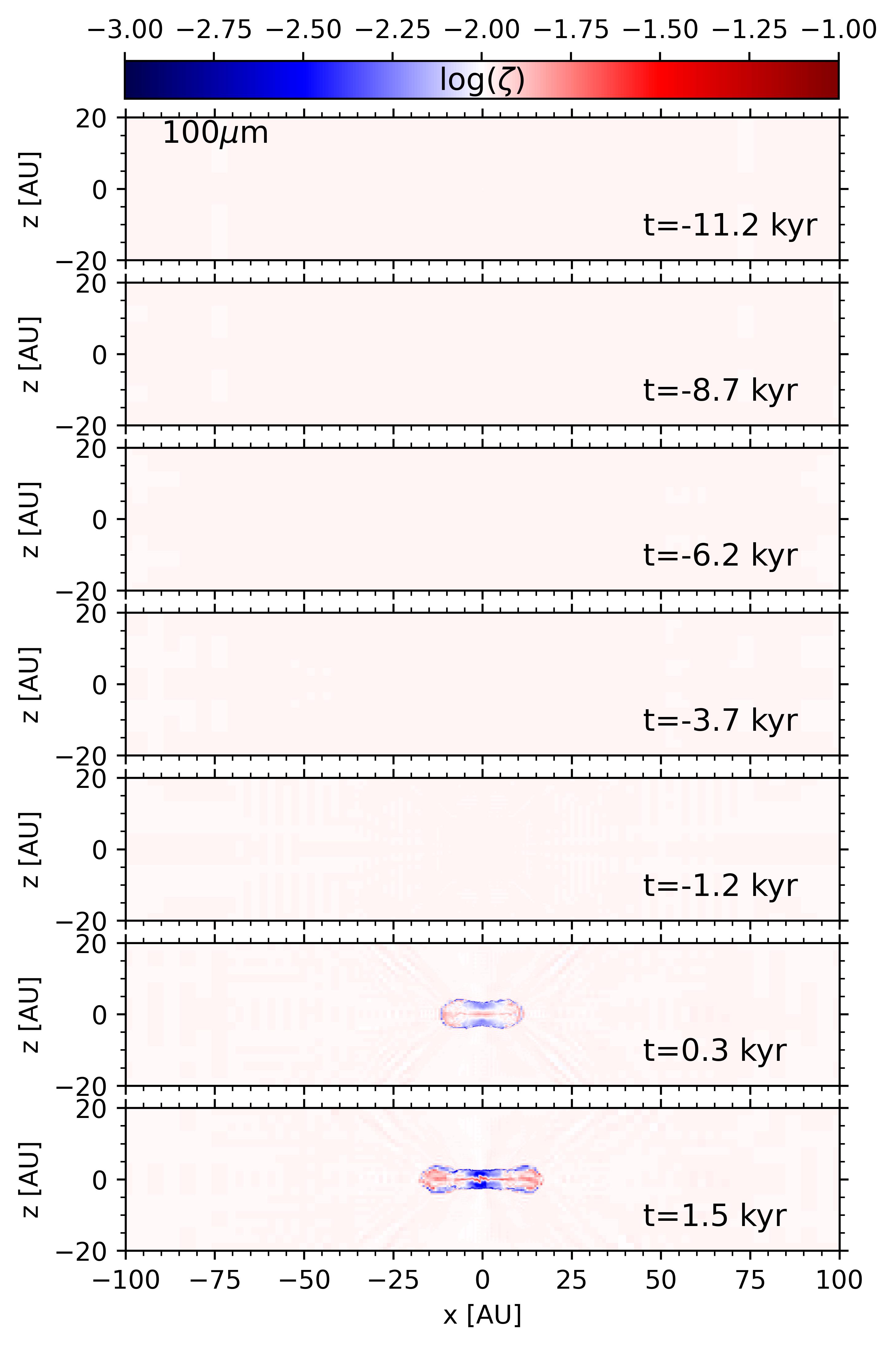}    
    \caption{ Time evolution of the grown dust-to-gas mass ratio $\zeta_{\rm gr}$  and the total dust-to-gas mass ratio $\zeta$ in the $x-z$ slice for the $a_{\rm max}^{\rm cap}=100$~$\mu$m model. The left and right groups of panels show $\zeta_{\rm gr}$ and  $\zeta$, respectively. White color indicates the initial value of $\log\zeta_{\rm gr}=-2.51$and shades of red correspond to dust enhancements. Time is measured from the disk formation epoch, which occurs 193~kyr after the onset of the simulations.}
    \label{fig:6plot_xz_dens_dust_5}
\end{figure*}

Figures~\ref{fig:2d_dens_dust_tot} and \ref{fig:2d_dens_dust_grown} show the dust-to-gas mass ratios in the disk midplane at the end of our simulations ($t=12$~kyr after disk formation). In particular, Fig.~\ref{fig:2d_dens_dust_tot} shows $\zeta_{\rm gr}=\rho_{\rm d,gr} / \rho_{\rm g}$, the ratio of the grown dust density to that of gas, while Fig.~\ref{fig:2d_dens_dust_grown} shows $\zeta=(\rho_{\rm d,gr} + \rho_{\rm d,sm}) / \rho_{\rm g}$, the ratio of the total dust density to that of gas. We note that the initial adopted values of $\zeta$ and $\zeta_{\rm gr}$ at the onset of the prestellar cloud core collapse are $\log{\zeta}=-2.0$ and $\log{\zeta_{\rm gr}}=-2.51$. 

All models show dust enhancements in the disk midplane across most of the disk. The black contour curves outline the disk as determined by the disk tracking algorithm (see Sect.~\ref{Sect:disk-track}).  The magnitude of this effect is stronger for models with larger allowed dust sizes, but even models with the lowest $a_{\rm max}^{\rm cap}=10$~$\mu$m and $a_{\rm max}^{\rm cap}=100$~$\mu$m  show notable dust enhancements in the disk midplane. Dust enhancement in these models may be due to a particular arrangement of Stokes numbers that increase both with radial distance and distance from the disk midplane (see Figs.~\ref{fig:1d_rad_all_mod_3} and \ref{fig:stokesVSamax}). This can cause a traffic jam effect in the disk, as dust particles at smaller distances drift more slowly. However, low values of the Stokes number in the disk midplane should reduce the efficiency of this effect on radial dust drift.  Vertical settling, however, may play an important role in dust enhancement, as demonstrated in Sect.~\ref{Sect:surf-dens}.

A closer look at the regions just outside the disk may provide another possible explanation for the observed dust enhancement. These regions are characterized by $\zeta$ and $\zeta_{\rm gr}$ values slightly higher than their counterparts at the onset of cloud core collapse ($\log{\zeta}=-1.98$ versus initial $\log{\zeta}=-\mathbf{2.0}$ and $\log{\zeta_{\rm gr}}=-2.49$ versus the initial $\log{\zeta_{\rm gr}}=-2.51$), suggesting that some dust enhancement may have occurred in the predisk collapse stage.

To test this hypothesis, we plot the ratios of $\zeta_{\rm gr}$ in the left column of Fig.~\ref{fig:6plot_xz_dens_dust_5}   for the $a_{\rm max}^{\rm cap}=100$~$\mu$m model during the evolution stage immediately preceding disk formation. Vertical slices through the $x-z$ plane at $y=0$ are shown for this case. Time is counted from the disk formation epoch, with negative time corresponding to the predisk stage.
At $t=-11.2$~kyr, well before the disk formation instance, the inner regions of the collapsing core are already enhanced in grown dust above the initial values of  $\log\zeta_{\rm gr}=-2.51$. At $t=-1.2$~kyr, the maximum grown dust enhancement in the center of the collapsing core reaches 
$\log\zeta_{\rm gr}=-2.09$ in the $a_{\rm max}^{\rm cap}=100$~$\mu$m model. When the disk begins to form, it already carries the legacy of this early enrichment phase in grown dust, as can be seen in the bottom panels of Fig.~\ref{fig:6plot_xz_dens_dust_5}. The maximum grown dust enrichment there is $\log\zeta_{\rm gr}=-1.79$. Efficient settling to the disk midplane is also observed in this model.

As the right column in Figure~\ref{fig:6plot_xz_dens_dust_5} shows, dust enhancement in the predisk stage is much smaller when the total dust-to-gas ratio $\zeta$ (rather than $\zeta_{\rm gr}$) is considered.  In particular,  in the $a_{\rm max}^{\rm cap}=100~\mu$m model, the maximum value just before disk formation is $\log \zeta= -1.98$. This is only a minor increase compared to the initial value of $\log \zeta=-2.0$ at the onset of collapse. Notable enhancements in $\zeta$ near the disk midplane begin to occur only after the disk has formed. This implies that dust growth, followed by the conversion of small to grown dust (the $S$-term in Eqs.~\ref{eq:cont_dust_small} and \ref{eq:cont_dust_grown}), is primarily responsible for the enhancement in grown dust ($\zeta_{\rm gr}$) in the predisk stage. Dust drift plays only a minor role in this stage.

\begin{figure}
    \centering
    \includegraphics[width=1\columnwidth]{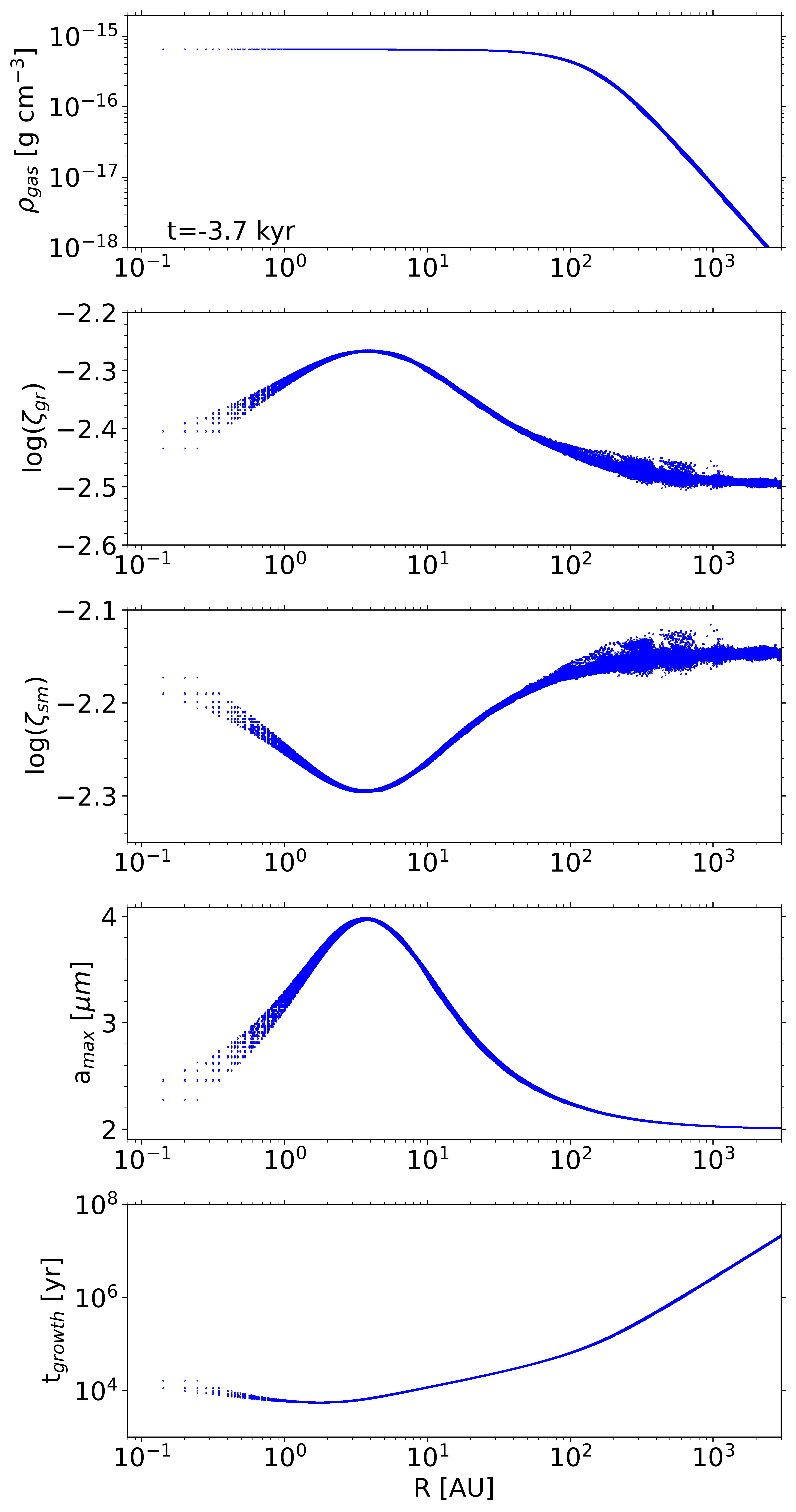}
    \caption{ Radial profiles of the collapsing cloud in the $a_{\rm max}^{\rm cap}=10$~$\mu$m model at 3.7~kyr before disk formation. From top to bottom: gas volume density, grown dust-to-gas mass ratio, small dust-to-gas mass ratio, maximum dust size, and dust growth timescale.}
    \label{fig:growth-source}
\end{figure}

We confirmed this conjecture by analyzing the radial distributions of $a_{\rm max}$, $\zeta_{\rm g}$, $\zeta - \zeta_{\rm gr}$ (i.e., the small dust-to-gas mass ratio) and the dust growth timescale during the predisk stage of cloud collapse. These distributions are shown in Fig.~\ref{fig:growth-source} for the $a_{\rm max}^{\rm cap}=10$~$\mu$m model at 3.7~kyr before disk formation. The gas density at the center of the collapsing cloud has increased by a factor of 500 compared to the initial distribution.
Dust growth predominantly occurs in this inner region because
the Stokes number reaches (and surpasses) unity there (see Fig.~\ref{fig:6plot_xz_dens_dust_10} in Appendix~\ref{App:three}). The dust growth timescale, calculated as $t_{\rm dust.growth} = a_{\rm max} /{\cal D}$, is about 10~kyr, much shorter than the collapse timescale, $t_{\rm collapse} = 193$~kyr. However, we note that the Stokes number is normalized to the Keplerian frequency $\Omega_{\rm K}$ in the growth rate term. This is relevant for circumstellar disks, where the turnover frequency of the largest turbulent eddy is typically assumed to equal the epicyclic frequency due to the Coriolis force, namely $\Omega_{\rm K}$ for Keplerian disks \citep{2007OrmelCuzzi}.  In the envelope, 
the eddy turnover frequency may differ; this case is discussed in more detail in Appendix~\ref{App:four}.  

.

\begin{figure}
    \centering
    \includegraphics[width=1\columnwidth]{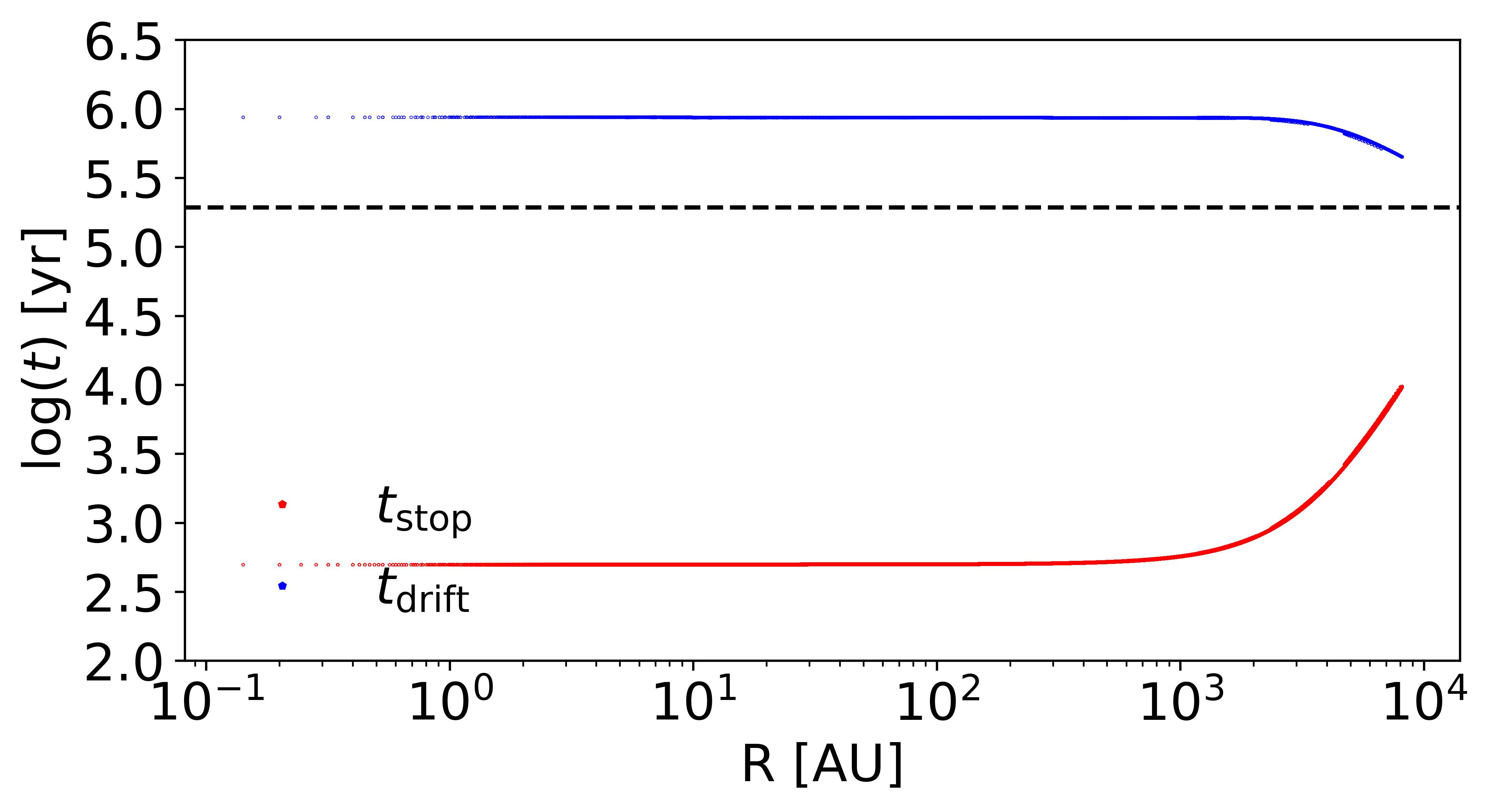}
    \caption{Characteristic dust drift time ($t_{\rm drift}$) and stopping  time ($t_{\rm stop}$), as functions of radial distance from the cloud core center. Both values are calculated at the onset of cloud core collapse. The horizontal dashed line indicates the time of FHSC formation.}
    \label{fig:timescales}
\end{figure}

\begin{figure}
    \centering
    \includegraphics[width=1\columnwidth]{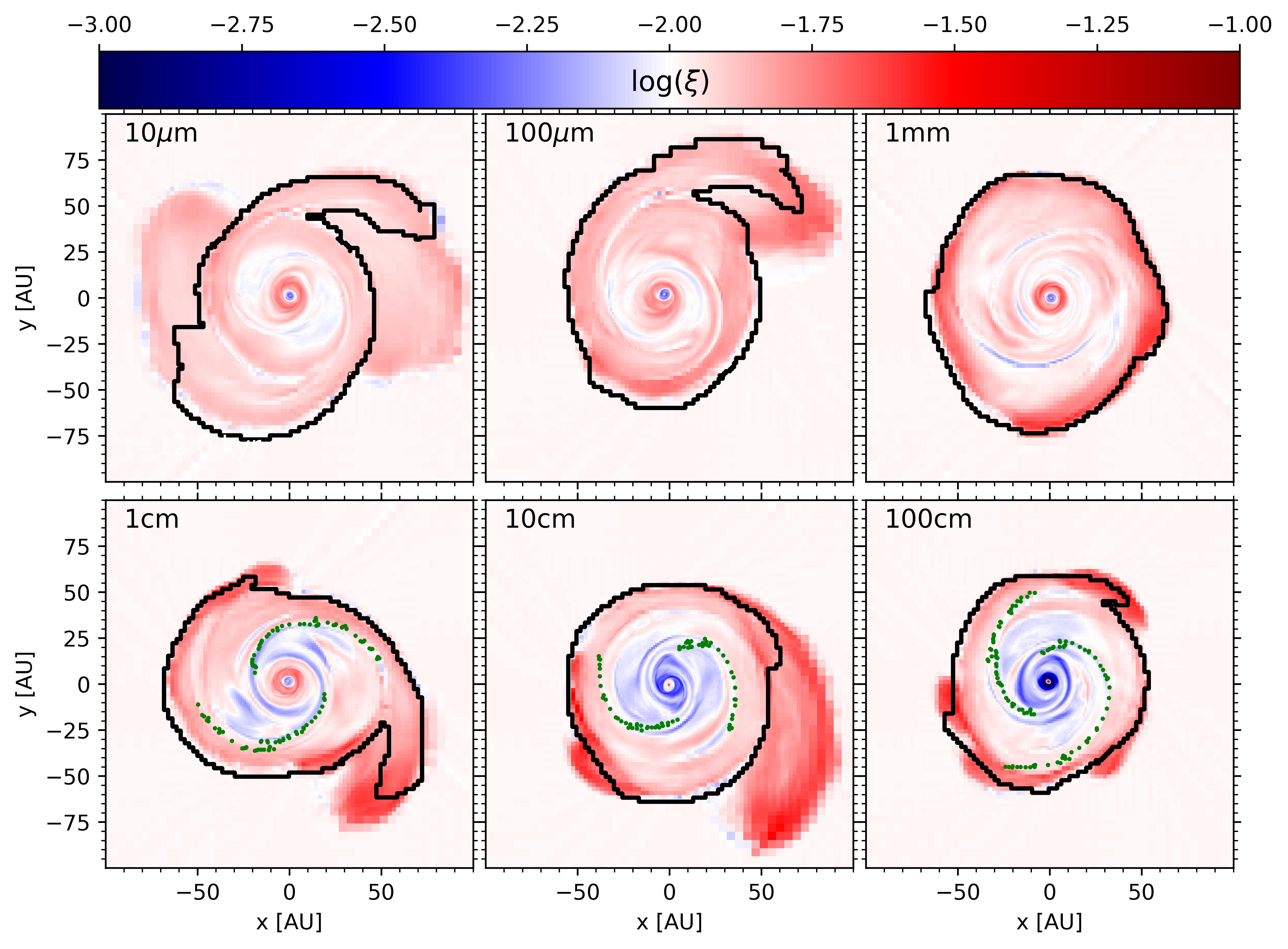}
    \caption{ Ratio of total dust to gas surface densities $\xi$, similar to Fig.~\ref{fig:2d_dens_dust_tot}. }
    \label{fig:xi-total}
\end{figure}

Figures~\ref{fig:6plot_xz_dens_dust_5}--\ref{fig:growth-source} demonstrate that dust growth and conversion of small to grown dust are responsible for the predisk enhancement in grown dust. The differential collapse of gas and dust in the envelope appears to play only a minor role.
We can further estimate the extent to which dust motion in the envelope is decoupled from gas motion by considering the initial stages of prestellar cloud collapse. Figure~\ref{fig:timescales} shows the stopping time of grown dust in the envelope at the onset of collapse, ranging from $<1.0$~kyr in the inner plateau to $\approx 10$~kyr at the cloud's outer edge. The time from the onset of collapse to FHSC formation is approximately $193$~yr, most of which is spent in slow cloud contraction. This means that the predisk phase of the gravitationally contracting cloud is much longer than the stopping time, implying that the infalling grown dust quickly attains equilibrium with the slowly contracting gas via mutual friction.

This state of equilibrium can be described by the relative velocity of grown dust to gas \citep{Birnstiel2024}:
\begin{equation}
    \bl{u} - \bl{v} = {t_{\rm stop} \over \rho_{\rm g} (1+\zeta_{\rm gr})} \bl{\nabla} P. 
\end{equation}
In this equation, the gravity force cancels out.
We can further calculate the drift timescale of the grown dust while it contracts with the gas as $t_{\rm drift}= R/|\bl{u} - \bl{v}|$, where $R$ is the radial distance from the center of the cloud. The corresponding values are plotted in Fig.~\ref{fig:timescales}, and are a factor of several times longer than the collapse timescale.  This suggests that dust enrichment due to drift in the predisk stage is insignificant, as confirmed by examining the total dust-to-gas ratios in the right panel of  Fig.~\ref{fig:6plot_xz_dens_dust_5}.

Finally, Figs.~\ref{fig:2d_dens_dust_tot} and \ref{fig:2d_dens_dust_grown} show the positions of the spiral arms, defined as local peaks in the midplane gas density, for the three models with the largest allowed dust sizes. Dust enhancements do not show a clear correlation with the positions of the spiral arms.  Several previous studies reported a correlation \citep{2004RiceLodato,Boss2020,2024MNRAS.528.2490R}, and this disagreement can be understood by comparing the Stokes numbers. In the aforementioned studies, dust efficiently concentrates in spiral arms when its Stokes number is around unity. However, such large Stokes numbers are rarely achievable in the earliest stages of disk formation. 
As described in \citet{VorobyovKulikov2024}, the disk at this stage is warmer and denser than its class II counterpart, and the corresponding $\mathrm{St}$ rarely exceeds 0.1 (see Fig.~\ref{fig:1d_rad_all_mod_3}).  The agile and nonsteady nature of the spiral pattern during this early evolutionary stage (see Fig.~\ref{fig:6x6_gas}) may also prevent clear dust concentration toward the spiral arms.

\subsection{Ratios of dust-to-gas surface densities}
\label{Sect:surf-dens}
In the previous section, we considered dust enhancement in the disk midplane. However, it is also important to consider the dust-to-gas mass ratios derived from the vertically integrated dust and gas volume densities, because the dust density in the disk midplane depends on disk turbulence, which remains unconstrained \citep{2023Rosotti}.  The resulting ratios $\xi=(\Sigma_{\rm d,gr}+\Sigma_{\rm d,sm} )/ \Sigma_{\rm g}$  are plotted in Fig.~\ref{fig:xi-total} for all six models at $t=12$~kyr. The surface densities of gas ($\Sigma_{\rm g}$), grown dust ($\Sigma_{\rm d,gr}$), and small dust ($\Sigma_{\rm d,sm}$) were calculated by vertically integrating the corresponding volume densities within a $100\times100\times100$~au cube centered on the coordinate center and encompassing the entire disk.

The ratio $\xi$ is highly nonhomogeneous across the disk. Local enhancements in dust alternate with depletions. Variations in $\xi$ around the initial value of 0.01 increase for higher $a_{\rm max}^{\rm cap}$ models.   
We calculated the total masses of dust (both small and grown) and gas within the $100\times100\times100$~au cube at $t=12$~kyr, finding an integrated dust-to-gas mass ratio of about 0.0104. This small positive deviation from the initial value of 0.01 is consistent with the previously established minor role of differential collapse of gas and dust in creating dust enhancements. The substantial local variations in $\xi$ seen in Figure~\ref{fig:xi-total} are likely due to local processes within the disk, such as the radially varying efficiency of vertical dust settling and perturbations by spiral density waves. More focused simulations are required to better understand the origin of these variations.

\section{Caveats}
\label{Sect:caveats}

Here, we discuss several model limitations that may affect dust enrichment and growth in the early stages of disk formation. Our simulations begin with the gravitational collapse of the Bonnor-Ebert sphere. If, instead, the initial conditions corresponded to a homogeneous sphere, the collapse time would likely be shorter, and the predisk dust enrichment weaker. Magnetic fields may also alter the character of cloud core collapse, influencing initial dust enrichment \citep{MachidaBasu2020,2024A&A...690A..23V}. The process of protostar formation during the second collapse of the FHSC is not included and may lead to a subtle interplay between the disk forming around the protostar and that forming around the FHSC \citep{Tomida2015}, potentially affecting dust dynamics and growth efficiency.  Finally, modeling dust growth in the collapsing envelope requires a different approach from that used in the disk \citep{Akimkin2022}.  All of these issues warrant further investigation.

\section{Conclusions}
\label{Sect:conclude}
In this work, we studied the initial stages of disk formation and evolution using three-dimensional numerical hydrodynamics simulations in the polytropic limit. The gas and dust dynamics were computed from the onset of the cloud core collapse to 12~kyr after FHSC formation. The simulations were terminated just before the collapse of the FHSC and the formation of the protostar.

Unlike our previous study \citep{VorobyovKulikov2024}, in which dust growth was halted at the fragmentation barrier corresponding to $v_{\rm frag}=5.0$~m~s$^{-1}$, here we adopted a range of $v_{\rm frag}$ from 0.5 to 50~m~s$^{-1}$, as suggested by various laboratory experiments, and also considered the effects of electrostatic and bouncing barriers.
This is implicitly done by setting several upper limits on the dust size $a_{\rm max}^{\rm cap}$ to which dust can grow, ranging from 10~$\mu$m to 100~cm, in multiples of ten. 
The lowest value of $a_{\rm max}^{\rm cap}$ can be viewed as that imposed by the electrostatic barrier, while the higher values of $a_{\rm max}^{\rm cap}$ correspond to various fragmentation (as defined by $v_{\rm frag}$) and bouncing barriers. The drift barrier to dust growth is accounted for self-consistently through the solution of the dust dynamics equation. Our findings are summarized as follows.

\begin{itemize}

\item  Dust growth begins during the collapse stage. Both the FHSC and the disk form in an environment already enriched in grown dust compared to the maximum dust size in the initial cloud core. The differential collapse of gas and dust in the cloud has only a minor effect on dust enrichment at the onset of disk formation.

\item The efficiency of dust growth during the disk formation stage is limited by dust growth barriers.
If the maximum dust grain size in a young protoplanetary disk is below 100~$\mu$m, dust growth is likely limited by bouncing or electrostatic barriers. For larger grains, collisional fragmentation and drift are the main dust growth barriers.

\item The disk midplane becomes quickly enriched with dust due to efficient dust settling. On the other hand, the vertically integrated dust distribution shows notable local deviations from the canonical 1:100 dust-to-gas mass ratio, whereas the globally integrated value deviates negligibly from the initial 1:100 value. Local hydrodynamic flows in a gravitationally unstable disk may account for these local variations, but the exact cause remains to be determined. 

\item In the fragmentation- and drift-limited regimes, the maximum dust size exhibits a profound negative radial gradient throughout most of the disk extend during the very early stages of evolution. A constant-dust-size approximation may not be appropriate in these cases. 

\item When choosing between the constant-size or constant-$\mathrm{St}$ approximations, the latter is preferable, particularly for disk conditions where dust can grow above 10~cm in size.

\end{itemize}

\begin{acknowledgements}  
We are grateful to the referee for an insightful review, which helped us to reconsider our initial conclusions and to improve the manuscript. We also grateful to the linguistic editor whose work greatly improved the readability of the manuscript.
 This work was supported by the FWF project I4311-N27 (E. I. V.) and RFBR project 19-51-14002 (A. S.).
Simulations were performed on the Vienna Scientific Cluster (\href{https://vsc.ac.at/}{https://vsc.ac.at/}) and on the Narval Cluster provided by Calcul Québec 
 (\href{https://www.calculquebec.ca/}{https://www.calculquebec.ca/}) and the Digital Research Alliance of Canada (\href{https://alliancecan.ca/}{https://alliancecan.ca/}).
\end{acknowledgements}

\bibliographystyle{aa}
\bibliography{refs}

\begin{thebibliography}{79}
\expandafter\ifx\csname natexlab\endcsname\relax\def\natexlab#1{#1}\fi

\bibitem[{{Ahmad} {et~al.}(2023){Ahmad}, {Gonz{\'a}lez}, {Hennebelle}, \& {Commer{\c{c}}on}}]{2023A&A...680A..23A}
{Ahmad}, A., {Gonz{\'a}lez}, M., {Hennebelle}, P., \& {Commer{\c{c}}on}, B. 2023, \aap, 680, A23

\bibitem[{{Akimkin} {et~al.}(2023{\natexlab{a}}){Akimkin}, {Ivlev}, {Caselli}, {Gong}, \& {Silsbee}}]{2023ApJ...953...72A}
{Akimkin}, V., {Ivlev}, A.~V., {Caselli}, P., {Gong}, M., \& {Silsbee}, K. 2023{\natexlab{a}}, \apj, 953, 72

\bibitem[{{Akimkin} {et~al.}(2023{\natexlab{b}}){Akimkin}, {Ivlev}, {Caselli}, {Gong}, \& {Silsbee}}]{2023Akimkin}
{Akimkin}, V., {Ivlev}, A.~V., {Caselli}, P., {Gong}, M., \& {Silsbee}, K. 2023{\natexlab{b}}, \apj, 953, 72

\bibitem[{{Akimkin}(2015)}]{Akimkin2015}
{Akimkin}, V.~V. 2015, Astronomy Reports, 59, 747

\bibitem[{{Akimkin} {et~al.}(2020){Akimkin}, {Ivlev}, \& {Caselli}}]{Akimkin2020_charge}
{Akimkin}, V.~V., {Ivlev}, A.~V., \& {Caselli}, P. 2020, \apj, 889, 64

\bibitem[{{ALMA Partnership} {et~al.}(2015){ALMA Partnership}, {Brogan}, {P{\'e}rez}, {Hunter}, {Dent}, {Hales}, {Hills}, {Corder}, {Fomalont}, {Vlahakis}, {Asaki}, {Barkats}, {Hirota}, {Hodge}, {Impellizzeri}, {Kneissl}, {Liuzzo}, {Lucas}, {Marcelino}, {Matsushita}, {Nakanishi}, {Phillips}, {Richards}, {Toledo}, {Aladro}, {Broguiere}, {Cortes}, {Cortes}, {Espada}, {Galarza}, {Garcia-Appadoo}, {Guzman-Ramirez}, {Humphreys}, {Jung}, {Kameno}, {Laing}, {Leon}, {Marconi}, {Mignano}, {Nikolic}, {Nyman}, {Radiszcz}, {Remijan}, {Rod{\'o}n}, {Sawada}, {Takahashi}, {Tilanus}, {Vila Vilaro}, {Watson}, {Wiklind}, {Akiyama}, {Chapillon}, {de Gregorio-Monsalvo}, {Di Francesco}, {Gueth}, {Kawamura}, {Lee}, {Nguyen Luong}, {Mangum}, {Pietu}, {Sanhueza}, {Saigo}, {Takakuwa}, {Ubach}, {van Kempen}, {Wootten}, {Castro-Carrizo}, {Francke}, {Gallardo}, {Garcia}, {Gonzalez}, {Hill}, {Kaminski}, {Kurono}, {Liu}, {Lopez}, {Morales}, {Plarre}, {Schieven}, {Testi}, {Videla}, {Villard}, {Andreani}, {Hibbard}, \&
  {Tatematsu}}]{2015ALMABrogan}
{ALMA Partnership}, {Brogan}, C.~L., {P{\'e}rez}, L.~M., {et~al.} 2015, \apjl, 808, L3

\bibitem[{{Bate}(2022)}]{Bate2022}
{Bate}, M.~R. 2022, \mnras, 514, 2145

\bibitem[{{Bhandare} {et~al.}(2020){Bhandare}, {Kuiper}, {Henning}, {Fendt}, {Flock}, \& {Marleau}}]{Kuiper2020}
{Bhandare}, A., {Kuiper}, R., {Henning}, T., {et~al.} 2020, \aap, 638, A86

\bibitem[{{Bhandare} {et~al.}(2018){Bhandare}, {Kuiper}, {Henning}, {Fendt}, {Marleau}, \& {K{\"o}lligan}}]{Kuiper2018}
{Bhandare}, A., {Kuiper}, R., {Henning}, T., {et~al.} 2018, \aap, 618, A95

\bibitem[{Binney {et~al.}(1987)Binney, Tremaine, \& Ostriker}]{1987Binney}
Binney, J., Tremaine, S., \& Ostriker, J. 1987, Galactic Dynamics, Princeton series in astrophysics (Princeton University Press)

\bibitem[{{Birnstiel}(2024)}]{Birnstiel2024}
{Birnstiel}, T. 2024, \araa, 62, 157

\bibitem[{{Birnstiel} {et~al.}(2016){Birnstiel}, {Fang}, \& {Johansen}}]{2016Birnstiel}
{Birnstiel}, T., {Fang}, M., \& {Johansen}, A. 2016, \ssr, 205, 41

\bibitem[{{Birnstiel} {et~al.}(2012){Birnstiel}, {Klahr}, \& {Ercolano}}]{2012Birnstiel}
{Birnstiel}, T., {Klahr}, H., \& {Ercolano}, B. 2012, \aap, 539, A148

\bibitem[{{Blum}(2018)}]{Blum2018}
{Blum}, J. 2018, \ssr, 214, 52

\bibitem[{{Blum} \& {Wurm}(2008)}]{2008ARA&A..46...21B}
{Blum}, J. \& {Wurm}, G. 2008, \araa, 46, 21

\bibitem[{{Boss} {et~al.}(2020){Boss}, {Alexander}, \& {Podolak}}]{Boss2020}
{Boss}, A.~P., {Alexander}, C. M.~O., \& {Podolak}, M. 2020, \apj, 901, 81

\bibitem[{{Cridland} {et~al.}(2022){Cridland}, {Rosotti}, {Tabone}, {Tychoniec}, {McClure}, {Nazari}, \& {van Dishoeck}}]{Cridland2022}
{Cridland}, A.~J., {Rosotti}, G.~P., {Tabone}, B., {et~al.} 2022, \aap, 662, A90

\bibitem[{{Dr{\k{a}}{\.z}kowska} {et~al.}(2023){Dr{\k{a}}{\.z}kowska}, {Bitsch}, {Lambrechts}, {Mulders}, {Harsono}, {Vazan}, {Liu}, {Ormel}, {Kretke}, \& {Morbidelli}}]{2023ASPC..534..717D}
{Dr{\k{a}}{\.z}kowska}, J., {Bitsch}, B., {Lambrechts}, M., {et~al.} 2023, in Astronomical Society of the Pacific Conference Series, Vol. 534, Protostars and Planets VII, ed. S.~{Inutsuka}, Y.~{Aikawa}, T.~{Muto}, K.~{Tomida}, \& M.~{Tamura}, 717

\bibitem[{{Galametz} {et~al.}(2019){Galametz}, {Maury}, {Valdivia}, {Testi}, {Belloche}, \& {Andr{\'e}}}]{2019Galametz}
{Galametz}, M., {Maury}, A.~J., {Valdivia}, V., {et~al.} 2019, \aap, 632, A5

\bibitem[{{Gundlach} \& {Blum}(2015)}]{2015ApJ...798...34G}
{Gundlach}, B. \& {Blum}, J. 2015, \apj, 798, 34

\bibitem[{{Hennebelle} {et~al.}(2020){Hennebelle}, {Commer{\c{c}}on}, {Lee}, \& {Charnoz}}]{2020Hennebelle}
{Hennebelle}, P., {Commer{\c{c}}on}, B., {Lee}, Y.-N., \& {Charnoz}, S. 2020, \aap, 635, A67

\bibitem[{{Houge} {et~al.}(2024){Houge}, {Mac{\'\i}as}, \& {Krijt}}]{Houge2024}
{Houge}, A., {Mac{\'\i}as}, E., \& {Krijt}, S. 2024, \mnras, 527, 9668

\bibitem[{{Inutsuka}(2012)}]{Inutsuka2012}
{Inutsuka}, S.-i. 2012, Progress of Theoretical and Experimental Physics, 2012, 01A307

\bibitem[{{Johansen} \& {Lambrechts}(2017)}]{2017JohansenLambrechts}
{Johansen}, A. \& {Lambrechts}, M. 2017, Annual Review of Earth and Planetary Sciences, 45, 359

\bibitem[{{Joos} {et~al.}(2012){Joos}, {Hennebelle}, \& {Ciardi}}]{2012Joos}
{Joos}, M., {Hennebelle}, P., \& {Ciardi}, A. 2012, \aap, 543, A128

\bibitem[{{Koga} \& {Machida}(2023)}]{2023Koga}
{Koga}, S. \& {Machida}, M.~N. 2023, \mnras, 519, 3595

\bibitem[{{Kratter} {et~al.}(2008){Kratter}, {Matzner}, \& {Krumholz}}]{2008Kratter}
{Kratter}, K.~M., {Matzner}, C.~D., \& {Krumholz}, M.~R. 2008, \apj, 681, 375

\bibitem[{{Laibe} \& {Price}(2014)}]{2014LaibePrice}
{Laibe}, G. \& {Price}, D.~J. 2014, \mnras, 444, 1940

\bibitem[{{Larson}(1985)}]{Larson1985}
{Larson}, R.~B. 1985, \mnras, 214, 379

\bibitem[{{Lebreuilly} {et~al.}(2020){Lebreuilly}, {Commer{\c{c}}on}, \& {Laibe}}]{2020Lebreuilly}
{Lebreuilly}, U., {Commer{\c{c}}on}, B., \& {Laibe}, G. 2020, \aap, 641, A112

\bibitem[{{Lee} {et~al.}(2020){Lee}, {Li}, \& {Turner}}]{Turner2020}
{Lee}, C.-F., {Li}, Z.-Y., \& {Turner}, N.~J. 2020, Nature Astronomy, 4, 142

\bibitem[{{Lin} \& {Youdin}(2017)}]{2017LinYoudin}
{Lin}, M.-K. \& {Youdin}, A.~N. 2017, \apj, 849, 129

\bibitem[{{Liu} {et~al.}(2021){Liu}, {Tsai}, {Chen}, {Liu}, {Zhang}, {Ma}, {Elbakyan}, {Green}, {Hales}, {Liu}, {Takami}, {P{\'e}rez}, {Vorobyov}, \& {Yang}}]{Liu2021}
{Liu}, H.~B., {Tsai}, A.-L., {Chen}, W.~P., {et~al.} 2021, \apj, 923, 270

\bibitem[{{Lor{\'e}n-Aguilar} \& {Bate}(2015)}]{2015LorenBate}
{Lor{\'e}n-Aguilar}, P. \& {Bate}, M.~R. 2015, \mnras, 454, 4114

\bibitem[{{Machida} \& {Basu}(2020)}]{MachidaBasu2020}
{Machida}, M.~N. \& {Basu}, S. 2020, \mnras, 494, 827

\bibitem[{{Machida} {et~al.}(2014){Machida}, {Inutsuka}, \& {Matsumoto}}]{2014Machida}
{Machida}, M.~N., {Inutsuka}, S.-i., \& {Matsumoto}, T. 2014, \mnras, 438, 2278

\bibitem[{{Machida} \& {Nakamura}(2015)}]{2015MachidaNakamura}
{Machida}, M.~N. \& {Nakamura}, T. 2015, \mnras, 448, 1405

\bibitem[{{Marchand} {et~al.}(2023){Marchand}, {Lebreuilly}, {Mac Low}, \& {Guillet}}]{Marchand2023}
{Marchand}, P., {Lebreuilly}, U., {Mac Low}, M.~M., \& {Guillet}, V. 2023, \aap, 670, A61

\bibitem[{{Masunaga} \& {Inutsuka}(2000{\natexlab{a}})}]{2000Masunaga}
{Masunaga}, H. \& {Inutsuka}, S.-i. 2000{\natexlab{a}}, \apj, 531, 350

\bibitem[{{Masunaga} \& {Inutsuka}(2000{\natexlab{b}})}]{Masunaga2000}
{Masunaga}, H. \& {Inutsuka}, S.-i. 2000{\natexlab{b}}, \apj, 531, 350

\bibitem[{{Mathis} {et~al.}(1977){Mathis}, {Rumpl}, \& {Nordsieck}}]{Mathis1977}
{Mathis}, J.~S., {Rumpl}, W., \& {Nordsieck}, K.~H. 1977, \apj, 217, 425

\bibitem[{{McKevitt} {et~al.}(2024){McKevitt}, {Vorobyov}, \& {Kulikov}}]{mcKevitt2024}
{McKevitt}, J., {Vorobyov}, E.~I., \& {Kulikov}, I. 2024, J. of Parallel and Distributed Comp., 195, 104977

\bibitem[{{Molyarova} {et~al.}(2021){Molyarova}, {Vorobyov}, {Akimkin}, {Skliarevskii}, {Wiebe}, \& {G{\"u}del}}]{Molyarova2021}
{Molyarova}, T., {Vorobyov}, E.~I., {Akimkin}, V., {et~al.} 2021, \apj, 910, 153

\bibitem[{{Morbidelli} {et~al.}(2024){Morbidelli}, {Marrocchi}, {Ali Ahmad}, {Bhandare}, {Charnoz}, {Commer{\c{c}}on}, {Dullemond}, {Guillot}, {Hennebelle}, {Lee}, {Lovascio}, {Marschall}, {Marty}, {Maury}, \& {Tamami}}]{Morbidelli2024}
{Morbidelli}, A., {Marrocchi}, Y., {Ali Ahmad}, A., {et~al.} 2024, \aap, 691, A147

\bibitem[{{Okuzumi}(2009)}]{2009Okuzumi}
{Okuzumi}, S. 2009, \apj, 698, 1122

\bibitem[{{Okuzumi} {et~al.}(2016){Okuzumi}, {Momose}, {Sirono}, {Kobayashi}, \& {Tanaka}}]{Okuzumi2016}
{Okuzumi}, S., {Momose}, M., {Sirono}, S.-i., {Kobayashi}, H., \& {Tanaka}, H. 2016, \apj, 821, 82

\bibitem[{{Okuzumi} {et~al.}(2011{\natexlab{a}}){Okuzumi}, {Tanaka}, {Takeuchi}, \& {Sakagami}}]{Okuzumi2011_charge}
{Okuzumi}, S., {Tanaka}, H., {Takeuchi}, T., \& {Sakagami}, M.-a. 2011{\natexlab{a}}, \apj, 731, 95

\bibitem[{{Okuzumi} {et~al.}(2011{\natexlab{b}}){Okuzumi}, {Tanaka}, {Takeuchi}, \& {Sakagami}}]{2011OkuzumiTanaka}
{Okuzumi}, S., {Tanaka}, H., {Takeuchi}, T., \& {Sakagami}, M.-a. 2011{\natexlab{b}}, \apj, 731, 95

\bibitem[{{Ormel} \& {Cuzzi}(2007)}]{2007OrmelCuzzi}
{Ormel}, C.~W. \& {Cuzzi}, J.~N. 2007, \aap, 466, 413

\bibitem[{{Pinte} {et~al.}(2016){Pinte}, {Dent}, {M{\'e}nard}, {Hales}, {Hill}, {Cortes}, \& {de Gregorio-Monsalvo}}]{Pinte2016}
{Pinte}, C., {Dent}, W.~R.~F., {M{\'e}nard}, F., {et~al.} 2016, \apj, 816, 25

\bibitem[{{Rice} {et~al.}(2004){Rice}, {Lodato}, {Pringle}, {Armitage}, \& {Bonnell}}]{2004RiceLodato}
{Rice}, W.~K.~M., {Lodato}, G., {Pringle}, J.~E., {Armitage}, P.~J., \& {Bonnell}, I.~A. 2004, \mnras, 355, 543

\bibitem[{{Rosotti}(2023)}]{2023Rosotti}
{Rosotti}, G.~P. 2023, \nar, 96, 101674

\bibitem[{{Rowther} {et~al.}(2024){Rowther}, {Nealon}, {Meru}, {Wurster}, {Aly}, {Alexander}, {Rice}, \& {Booth}}]{2024MNRAS.528.2490R}
{Rowther}, S., {Nealon}, R., {Meru}, F., {et~al.} 2024, \mnras, 528, 2490

\bibitem[{{Sai} {et~al.}(2023){Sai}, {Ohashi}, {Yen}, {Maury}, \& {Maret}}]{2023ApJ...944..222S}
{Sai}, J. I.~C., {Ohashi}, N., {Yen}, H.-W., {Maury}, A.~J., \& {Maret}, S. 2023, \apj, 944, 222

\bibitem[{{Sengupta} {et~al.}(2024){Sengupta}, {Cuzzi}, {Umurhan}, \& {Lyra}}]{2024ApJ...966...90S}
{Sengupta}, D., {Cuzzi}, J.~N., {Umurhan}, O.~M., \& {Lyra}, W. 2024, \apj, 966, 90

\bibitem[{{Silsbee} {et~al.}(2022){Silsbee}, {Akimkin}, {Ivlev}, {Testi}, {Gong}, \& {Caselli}}]{Akimkin2022}
{Silsbee}, K., {Akimkin}, V., {Ivlev}, A.~V., {et~al.} 2022, \apj, 940, 188

\bibitem[{{Steinpilz} {et~al.}(2020){Steinpilz}, {Joeris}, {Jungmann}, {Wolf}, {Brendel}, {Teiser}, {Shinbrot}, \& {Wurm}}]{2020Steinpilz}
{Steinpilz}, T., {Joeris}, K., {Jungmann}, F., {et~al.} 2020, Nature Physics, 16, 225

\bibitem[{{Stoyanovskaya} {et~al.}(2018){Stoyanovskaya}, {Vorobyov}, \& {Snytnikov}}]{2018Stoyanovskaya}
{Stoyanovskaya}, O.~P., {Vorobyov}, E.~I., \& {Snytnikov}, V.~N. 2018, Astronomy Reports, 62, 455

\bibitem[{{Tazzari} {et~al.}(2016){Tazzari}, {Testi}, {Ercolano}, {Natta}, {Isella}, {Chandler}, {P{\'e}rez}, {Andrews}, {Wilner}, {Ricci}, {Henning}, {Linz}, {Kwon}, {Corder}, {Dullemond}, {Carpenter}, {Sargent}, {Mundy}, {Storm}, {Calvet}, {Greaves}, {Lazio}, \& {Deller}}]{2016A&A...588A..53T}
{Tazzari}, M., {Testi}, L., {Ercolano}, B., {et~al.} 2016, \aap, 588, A53

\bibitem[{{Tomida} {et~al.}(2010){Tomida}, {Machida}, {Saigo}, {Tomisaka}, \& {Matsumoto}}]{Tomida2010}
{Tomida}, K., {Machida}, M.~N., {Saigo}, K., {Tomisaka}, K., \& {Matsumoto}, T. 2010, \apjl, 725, L239

\bibitem[{{Tomida} {et~al.}(2015){Tomida}, {Okuzumi}, \& {Machida}}]{Tomida2015}
{Tomida}, K., {Okuzumi}, S., \& {Machida}, M.~N. 2015, \apj, 801, 117

\bibitem[{Toro(2019)}]{Toro2019}
Toro, E. 2019, Shock Waves, 29, 1065

\bibitem[{{Tsukamoto} {et~al.}(2021){Tsukamoto}, {Machida}, \& {Inutsuka}}]{2021Tsukamoto}
{Tsukamoto}, Y., {Machida}, M.~N., \& {Inutsuka}, S.-i. 2021, \apjl, 920, L35

\bibitem[{{Tsukamoto} {et~al.}(2017){Tsukamoto}, {Okuzumi}, {Iwasaki}, {Machida}, \& {Inutsuka}}]{Tsukamoto2017}
{Tsukamoto}, Y., {Okuzumi}, S., {Iwasaki}, K., {Machida}, M.~N., \& {Inutsuka}, S.-i. 2017, \pasj, 69, 95

\bibitem[{{Valdivia} {et~al.}(2019){Valdivia}, {Maury}, {Brauer}, {Hennebelle}, {Galametz}, {Guillet}, \& {Reissl}}]{Valdivia2019}
{Valdivia}, V., {Maury}, A., {Brauer}, R., {et~al.} 2019, \mnras, 488, 4897

\bibitem[{{Vallucci-Goy} {et~al.}(2024){Vallucci-Goy}, {Lebreuilly}, \& {Hennebelle}}]{2024A&A...690A..23V}
{Vallucci-Goy}, V., {Lebreuilly}, U., \& {Hennebelle}, P. 2024, \aap, 690, A23

\bibitem[{{van der Marel} {et~al.}(2019){van der Marel}, {Dong}, {di Francesco}, {Williams}, \& {Tobin}}]{VanDerMarel2019}
{van der Marel}, N., {Dong}, R., {di Francesco}, J., {Williams}, J.~P., \& {Tobin}, J. 2019, \apj, 872, 112

\bibitem[{{Vorobyov} {et~al.}(2018){Vorobyov}, {Akimkin}, {Stoyanovskaya}, {Pavlyuchenkov}, \& {Liu}}]{2018VorobyovAkimkin}
{Vorobyov}, E.~I., {Akimkin}, V., {Stoyanovskaya}, O., {Pavlyuchenkov}, Y., \& {Liu}, H.~B. 2018, \aap, 614, A98

\bibitem[{{Vorobyov} \& {Basu}(2005)}]{VorobyovBasu2005}
{Vorobyov}, E.~I. \& {Basu}, S. 2005, \apjl, 633, L137

\bibitem[{{Vorobyov} {et~al.}(2023{\natexlab{a}}){Vorobyov}, {Elbakyan}, {Johansen}, {Lambrechts}, {Skliarevskii}, \& {Stoyanovskaya}}]{Vorobyov2023a}
{Vorobyov}, E.~I., {Elbakyan}, V.~G., {Johansen}, A., {et~al.} 2023{\natexlab{a}}, \aap, 670, A81

\bibitem[{{Vorobyov} {et~al.}(2024){Vorobyov}, {Kulikov}, {Elbakyan}, {McKevitt}, \& {G{\"u}del}}]{VorobyovKulikov2024}
{Vorobyov}, E.~I., {Kulikov}, I., {Elbakyan}, V.~G., {McKevitt}, J., \& {G{\"u}del}, M. 2024, \aap, 683, A202

\bibitem[{{Vorobyov} {et~al.}(2023{\natexlab{b}}){Vorobyov}, {McKevitt}, {Kulikov}, \& {Elbakyan}}]{VorobyovMcKevitt2023}
{Vorobyov}, E.~I., {McKevitt}, J., {Kulikov}, I., \& {Elbakyan}, V. 2023{\natexlab{b}}, \aap, 671, A81

\bibitem[{{Vorobyov} {et~al.}(2019){Vorobyov}, {Skliarevskii}, {Elbakyan}, {Pavlyuchenkov}, {Akimkin}, \& {Guedel}}]{2019VorobyovSkliarevskii}
{Vorobyov}, E.~I., {Skliarevskii}, A.~M., {Elbakyan}, V.~G., {et~al.} 2019, \aap, 627, A154

\bibitem[{{Vorobyov} {et~al.}(2022){Vorobyov}, {Skliarevskii}, {Molyarova}, {Akimkin}, {Pavlyuchenkov}, {K{\'o}sp{\'a}l}, {Liu}, {Takami}, \& {Topchieva}}]{Vorobyov2022}
{Vorobyov}, E.~I., {Skliarevskii}, A.~M., {Molyarova}, T., {et~al.} 2022, \aap, 658, A191

\bibitem[{{Wada} {et~al.}(2009){Wada}, {Tanaka}, {Suyama}, {Kimura}, \& {Yamamoto}}]{Wada2009}
{Wada}, K., {Tanaka}, H., {Suyama}, T., {Kimura}, H., \& {Yamamoto}, T. 2009, \apj, 702, 1490

\bibitem[{{Wada} {et~al.}(2011){Wada}, {Tanaka}, {Suyama}, {Kimura}, \& {Yamamoto}}]{Wada2011}
{Wada}, K., {Tanaka}, H., {Suyama}, T., {Kimura}, H., \& {Yamamoto}, T. 2011, \apj, 737, 36

\bibitem[{{Ward-Thompson} {et~al.}(2005){Ward-Thompson}, {Hartmann}, \& {Nutter}}]{2005MNRAS.357..687W}
{Ward-Thompson}, D., {Hartmann}, L., \& {Nutter}, D.~J. 2005, \mnras, 357, 687

\bibitem[{{Windmark} {et~al.}(2012){Windmark}, {Birnstiel}, {Ormel}, \& {Dullemond}}]{Windmark2012}
{Windmark}, F., {Birnstiel}, T., {Ormel}, C.~W., \& {Dullemond}, C.~P. 2012, \aap, 544, L16

\bibitem[{{Zsom} {et~al.}(2010){Zsom}, {Ormel}, {G{\"u}ttler}, {Blum}, \& {Dullemond}}]{2010ZsomOrmel}
{Zsom}, A., {Ormel}, C.~W., {G{\"u}ttler}, C., {Blum}, J., \& {Dullemond}, C.~P. 2010, \aap, 513, A57

\end{thebibliography}

\begin{appendix}
\section{Dust as a fluid}
\label{App:one}
It is not obvious that dist dynamics can be described using the gas hydrodynamic equations. One of the principle requirements is that the number of dust particles per computational cell is sufficiently large so that the mean dust densities and velocities obtained by integrating the  distribution function of dust particles over the velocity space do not suffer from statistical noise.  This condition is usually fulfilled with a large margin for a typical spectrum of dust sizes in a protoplanetary disk \citep[e.g.,][]{Vorobyov2022}. Another requirement is that the stopping length, defined as $\lambda_{\rm stop }= |\bl{u}-\bl{v}| t_{\rm stop} $, should be shorter than the size of the cell, by analogy to the requirement on the mean free path of a gas molecule (or atom) in gas dynamics. 

\begin{figure}
    \centering
    \includegraphics[width=1\columnwidth]{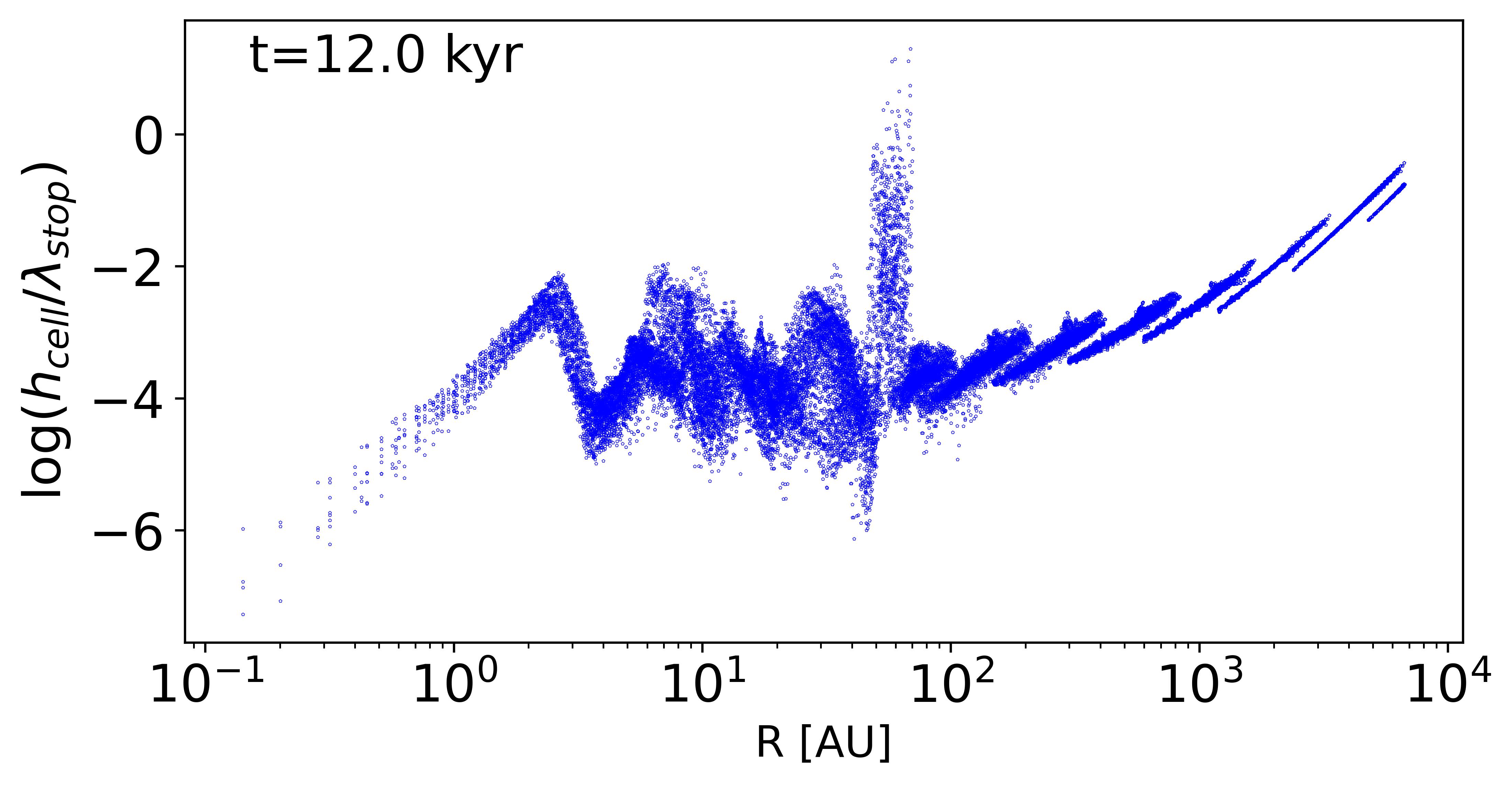}
    \caption{{Ratio of the cell size to the stopping length as a function of distance from the coordinate center.}}
    \label{fig:stopping_length}
\end{figure}

\begin{figure}
    \centering
    \includegraphics[width=0.47\columnwidth]{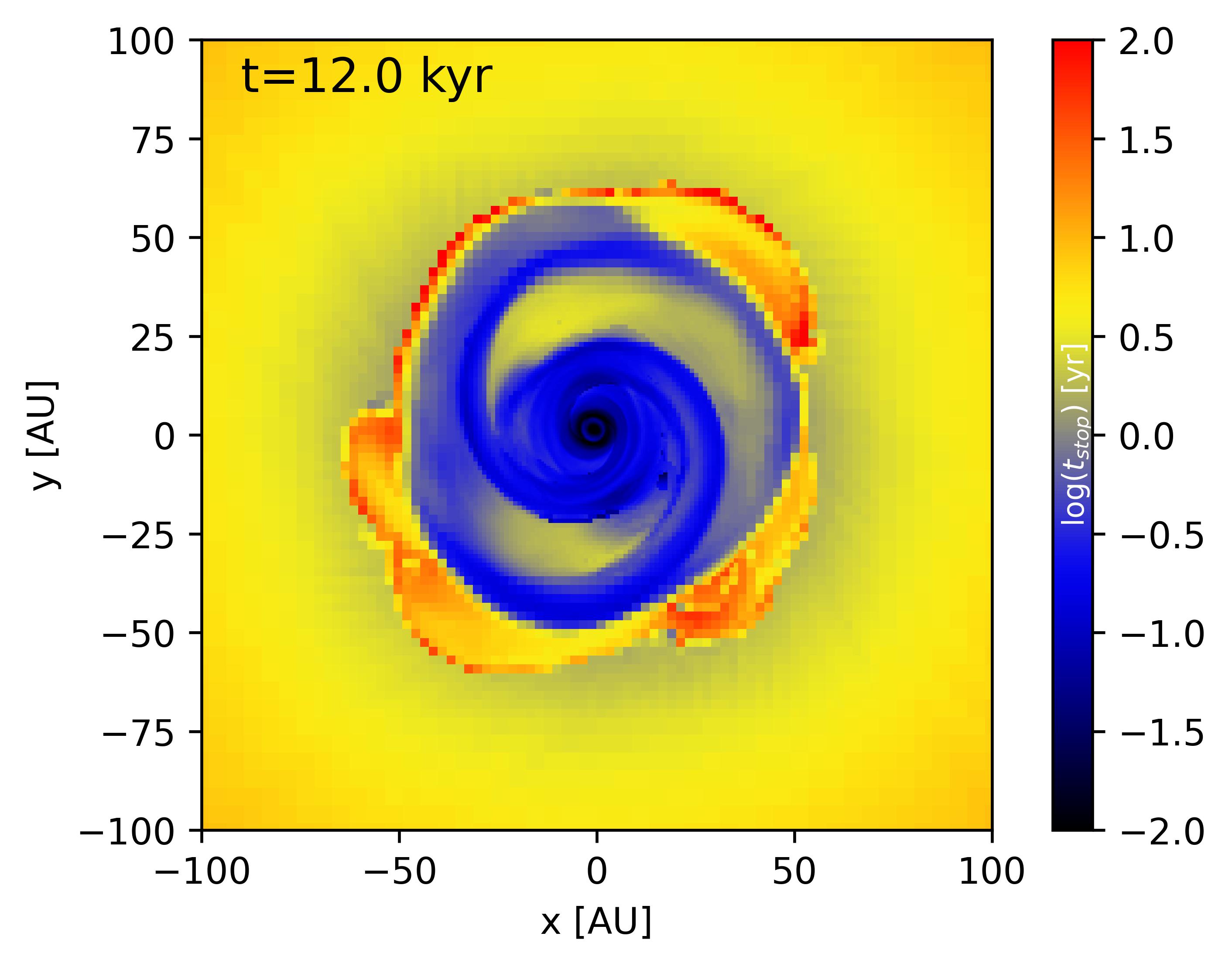}
    \includegraphics[width=0.47\columnwidth]{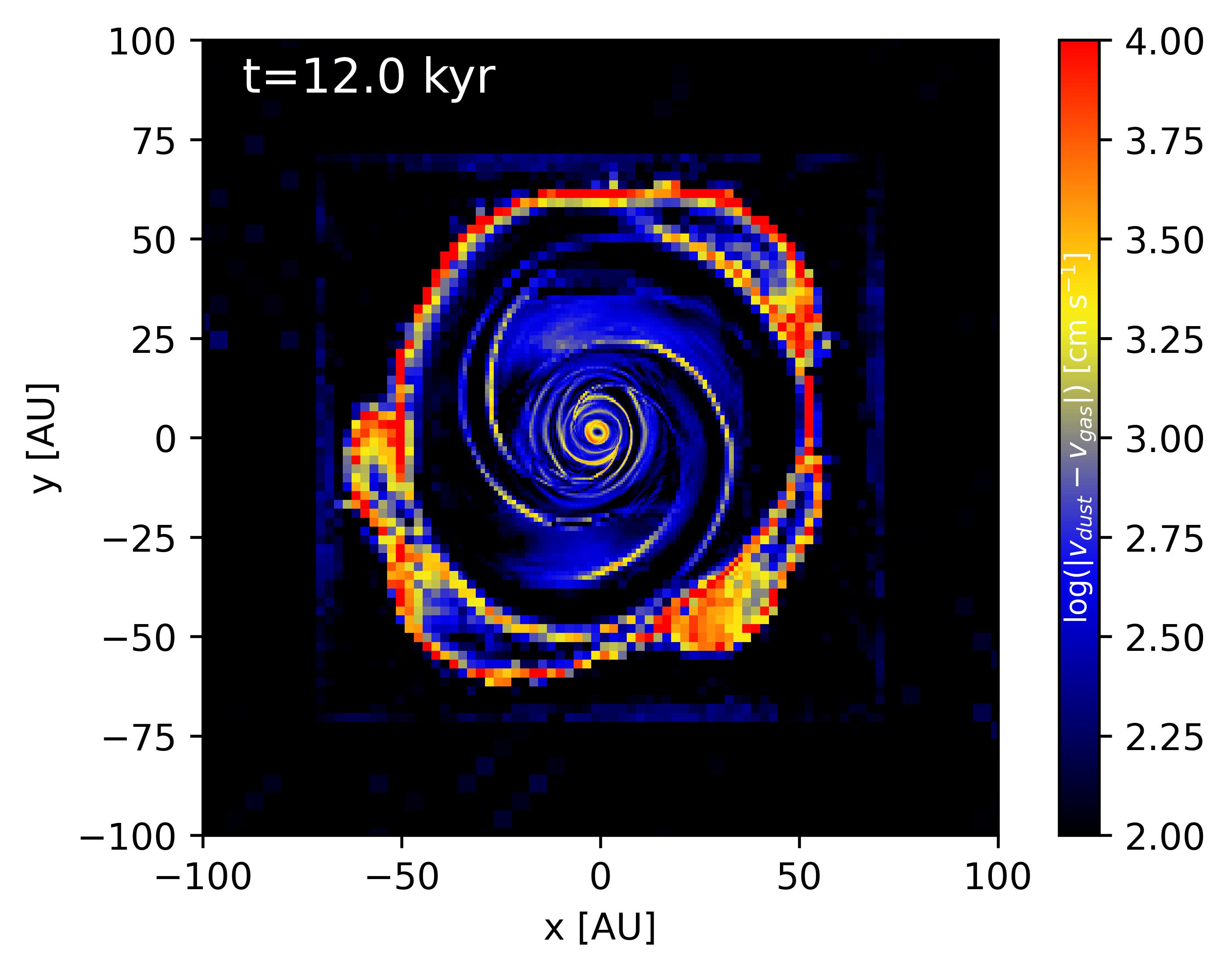}
    \caption{Stopping time (top panel) and the dust-minus-gas relative velocity by absolute value (bottom panel). The scale bars are in $\log$~yr and $\log$~cm~s$^{-1}$. }
    \label{fig:stopping_expl}
\end{figure}

This requirement is verified in Figure~\ref{fig:stopping_length}, which shows the ratio of the cell size to the stopping length as a function of distance from the coordinate center for model with $a_{\rm max}^{\rm cap}=100$~cm at $t=12$~kyr. This model is supposed to have the longest stopping time and, thus, the largest stopping length. The stopping length is much smaller than the size of the cell in the disk and in the collapsing envelope. However, there is one notable exception -- the interface between the disk and the envelope around 70~au where the ratio can exceed unity.  The cause can be seen in Figure~\ref{fig:stopping_expl} showing the two-dimensional spatial maps of $t_{\rm stop}$ and $|\bl{u}-\bl{v}|$ in the disk midplane. Both quantities are higher at the disk-envelope interface than in the disk and in the inner envelope. The higher values of the relative velocity are likely caused by the shock, which forms at the region where the infalling envelope meets the disk. The infalling gas is effectively decelerated at the shock while the pressureless dust reacts with a delay caused by the dust-to-gas friction. The stopping time is increased because dust experiences the first major episode of growth at the turbulent disk outer edge (see Fig.~\ref{fig:1d_rad_all_mod_2}). In anyway, this peculiar region is narrow and transitory, and thus should not seriously affect our modeling.

Finally, we note that a purely pressureless fluid is subject to the development of delta shocks, which lead to uncontrolled growth of density at discontinuities. It is important for numerical schemes to reproduce such a feature, as shown in Fig.~C.9 of \citet{VorobyovKulikov2024}. However, these features also corrupt the flow of a pressureless fluid and the fluid approach becomes inapplicable. Fortunately, delta shocks do not form if a repulsion force other than pressure is present in the system, and this is the case in our dust dynamics model where the role of the repulsion force is played by the gas-to-dust friction (see, the $\bl{f}$-term in Eq.~\ref{eq:mom-dust}). As Fig.~C.7 in \citet{VorobyovKulikov2024} demonstrates, the dust flow becomes smooth if dust to gas coupling is strong, correctly reproducing the shock and discontinuity positions. This is the case for the disk in which the Stokes numbers do not exceed 1.0 and are mostly below 0.1.

\section{Maximum dust size and Stokes number throughout the vertical extent of the disk}
\label{App:two}
\begin{figure*}
    \centering
    \includegraphics[width=0.65\columnwidth]{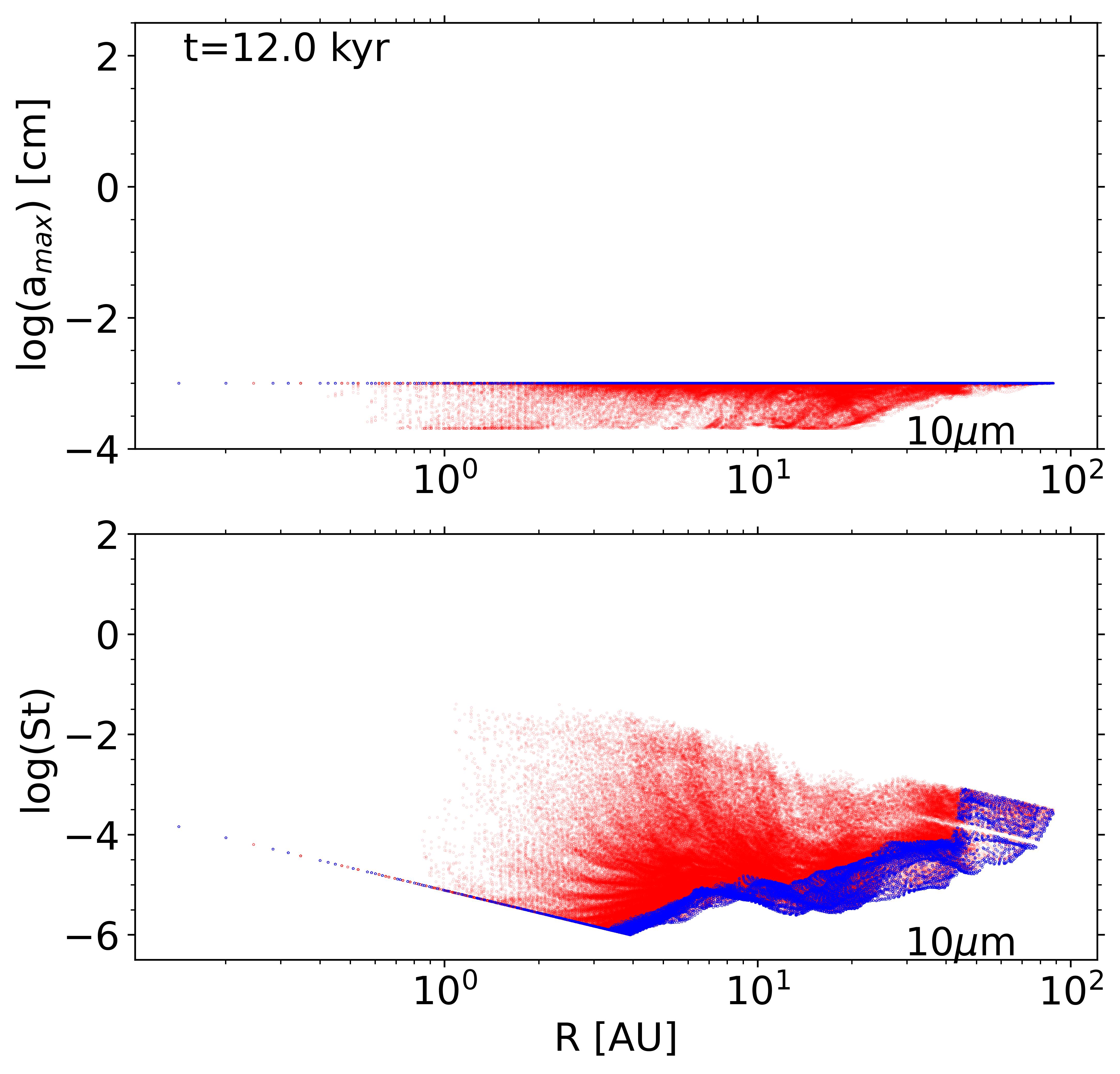}
    \includegraphics[width=0.65\columnwidth]{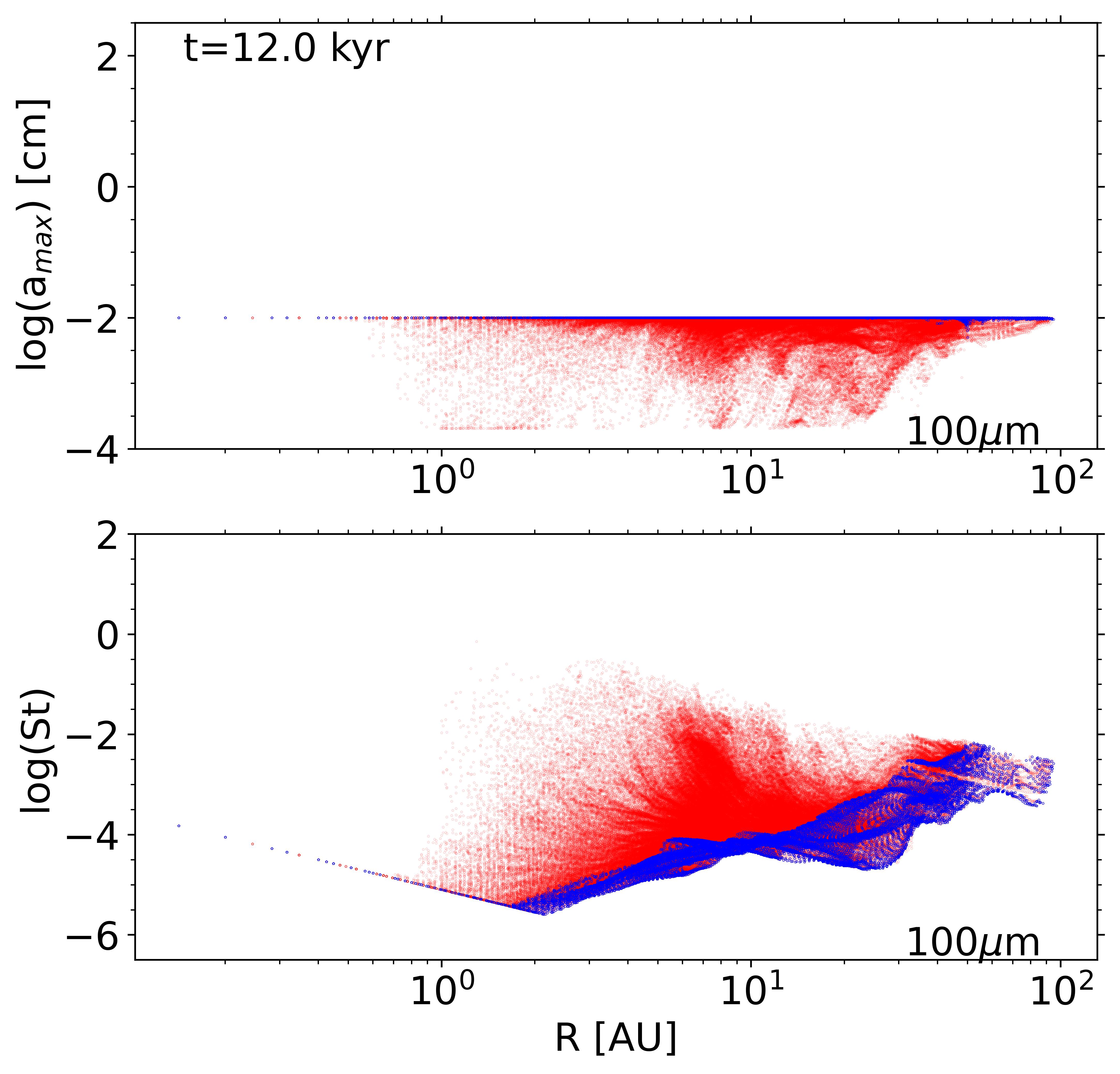}
    \includegraphics[width=0.65\columnwidth]{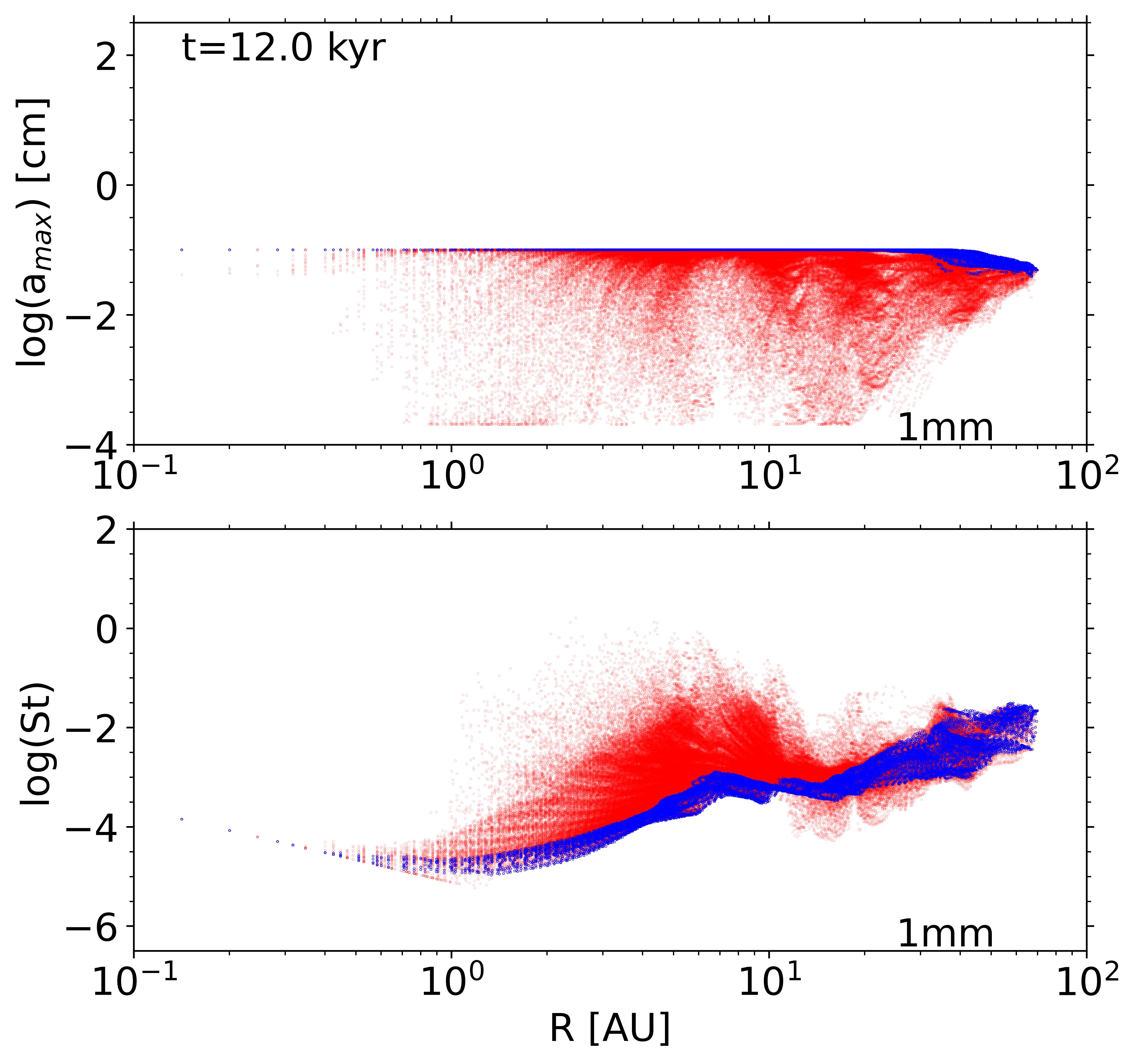}
    \includegraphics[width=0.65\columnwidth]{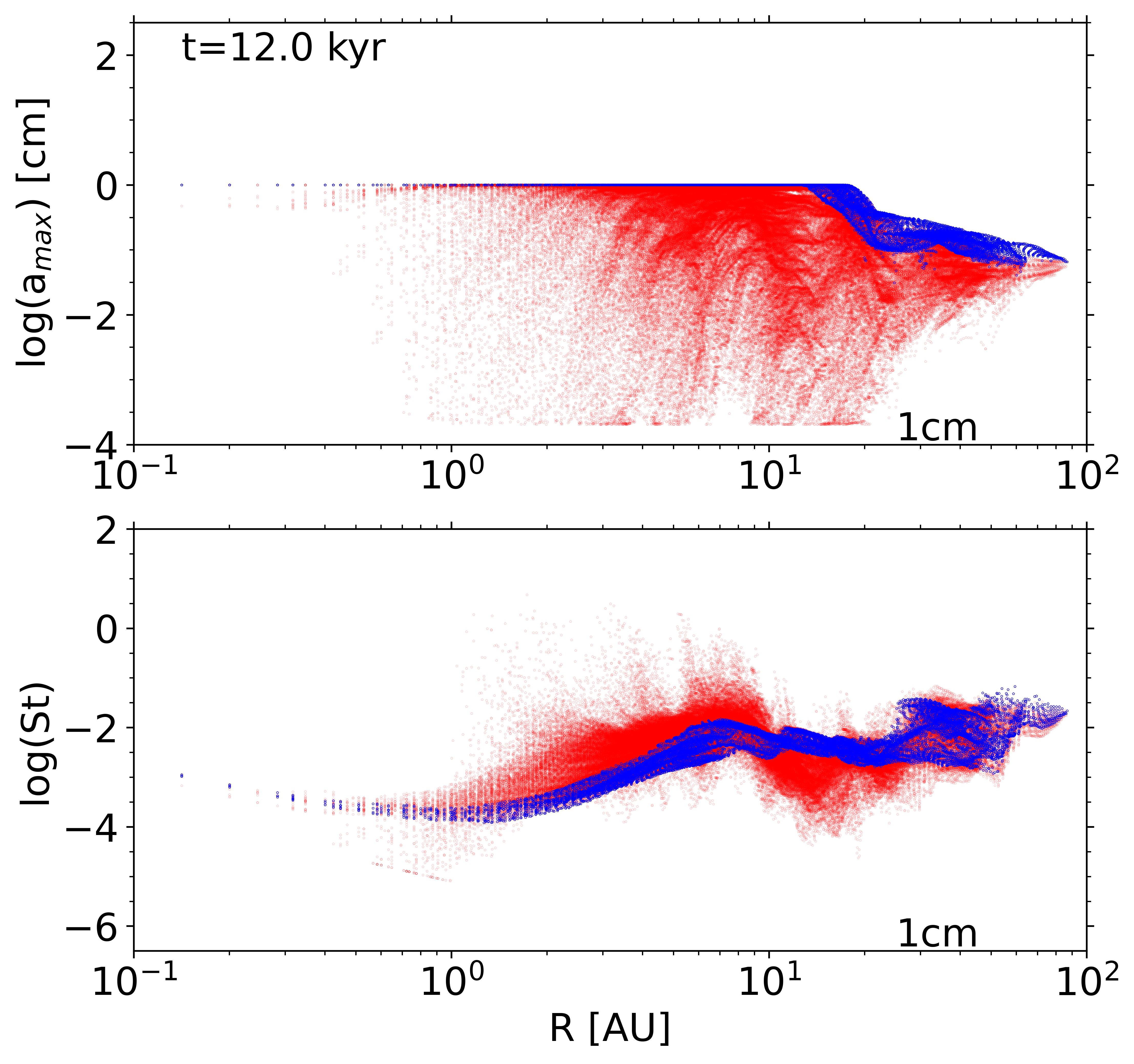}
    \includegraphics[width=0.65\columnwidth]{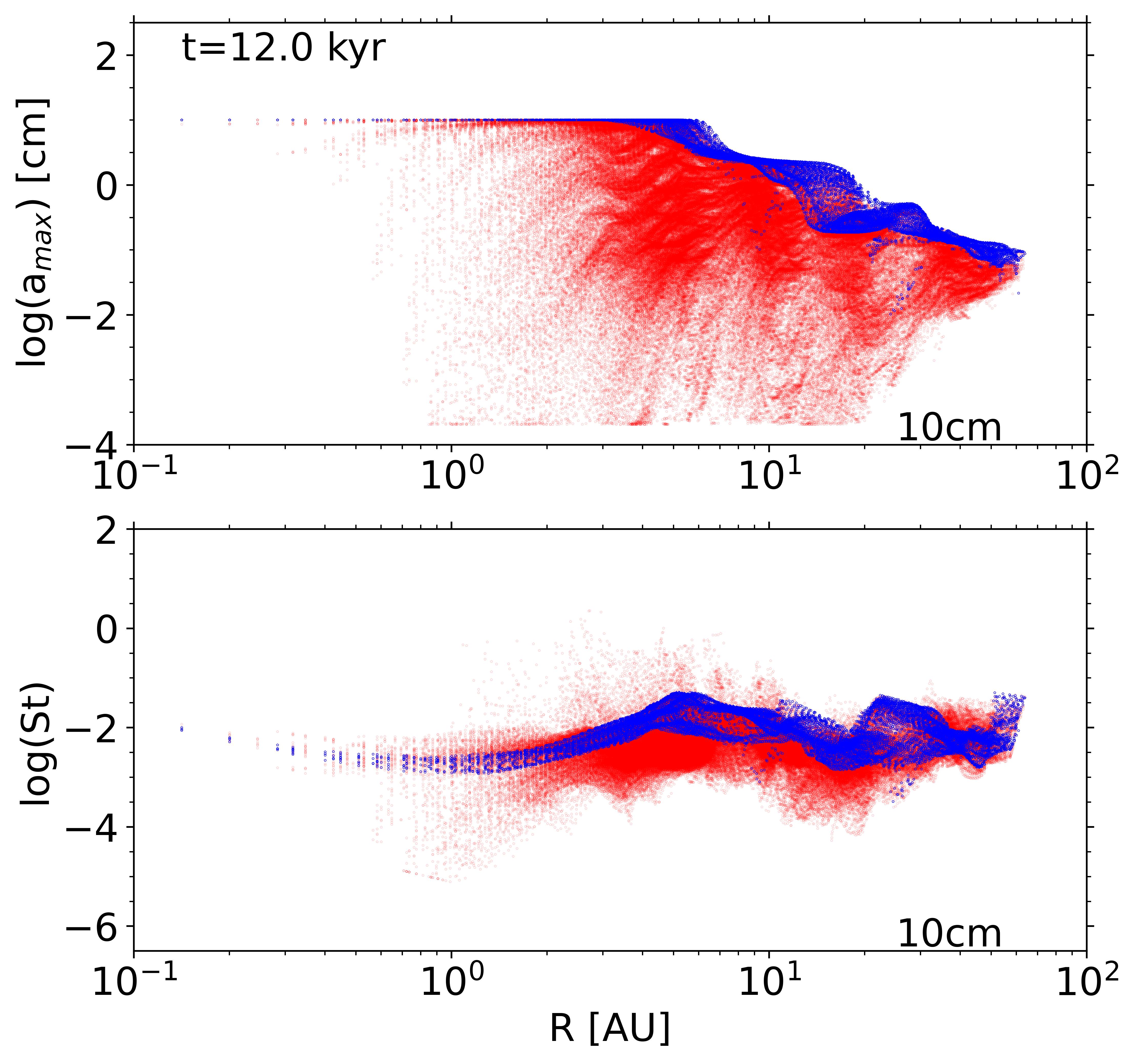}
    \includegraphics[width=0.65\columnwidth]{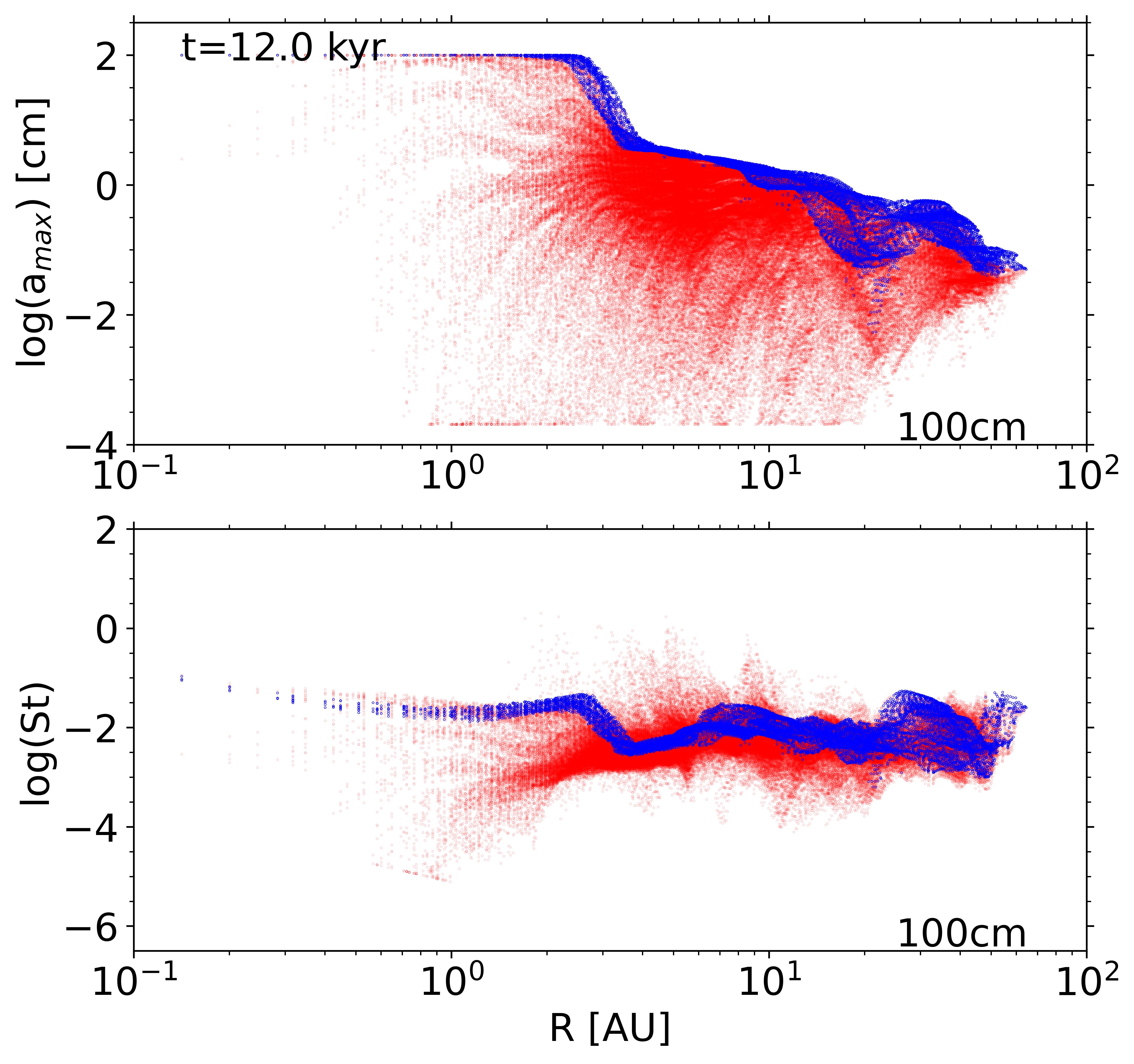}
    \caption{Dust maximum size and Stokes number in the disk midplane (blue dots) and disk atmosphere (red dots) as a function of distance in the disk midplane. The envelope is excluded. Two pairs of rows show, from left to right, the models with $a_{\rm max}^{\rm cap}$=10~$\mu$m, 100~$\mu$m, 1.0~mm, 1.0~cm, 10~cm, and 100~cm.}
    \label{fig:stokesVSamax}
\end{figure*}

Here, we briefly discuss the spatial distributions of the maximum dust size and Stokes number in the entire disk volume. Figure~\ref{fig:stokesVSamax} showa the corresponding values as a function of distance in the disk midplane for models with all six $a_{\rm max}^{\rm cap}$ considered. The blue dots represent the midplane values, while the red dots correspond to those in the rest of the disk (hereafter, disk atmosphere). Dust grains of maximum size are found in the disk midplane and the maximum dust size decreases in the disk atmosphere. The lowest values of $a_{\rm max}$ are found near the disk-envelope interface, approaching the values typical for the envelope, $a_{\rm max}=2.0$~$\mu$m. This distribution of dust grain sizes is expected for disks with vertical dust settling.

The Stokes number exhibits a more complicated trend. For models with $a_{\rm max}^{\rm cap}=10$~$\mu$m and 100~$\mu$m, the spatial distribution of $\mathrm{St}$ is opposite to that of $a_{\rm max}$ -- the lowest $\mathrm{St}$ are found in the disk midplane, while its values increase in the disk atmosphere.
For models with $a_{\rm max}^{\rm cap}\ge 1.0$~cm, this anticorrelation diminishes and the disk atmosphere is mostly characterized by $\mathrm{St}$ that is generally lower than that in the disk midplane. Considering that the disk conditions (gas densities and temperatures) do not change notably from model to model (see Fig.~\ref{fig:6x6_gas}), this difference in the Stokes number behavior can be attributed to efficient dust growth in the disk midplane. Although $\mathrm{St}$ may demonstrate an order of magnitude variations, models with a constant Stokes number are still preferable over those with a constant dust size when dust growth is allowed above 10~cm (e.g., models with $a_{\rm max}^{\rm cap}$=10 and 100~cm).

\section{Stokes number in the infalling envelope of the predisk stage}
\label{App:three}
\begin{figure*}
    \centering
    \includegraphics[width=1\columnwidth]{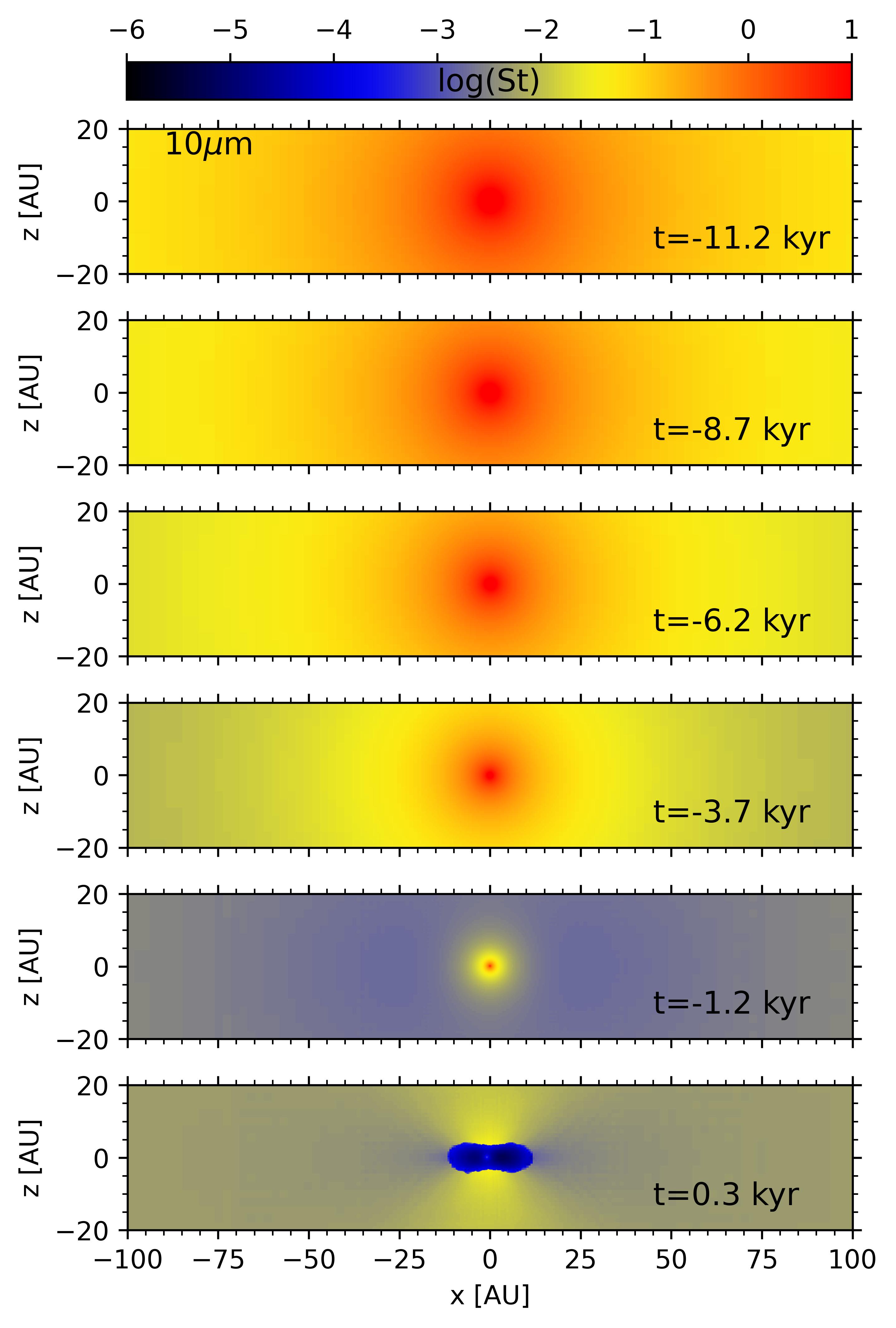}
    \includegraphics[width=1\columnwidth]{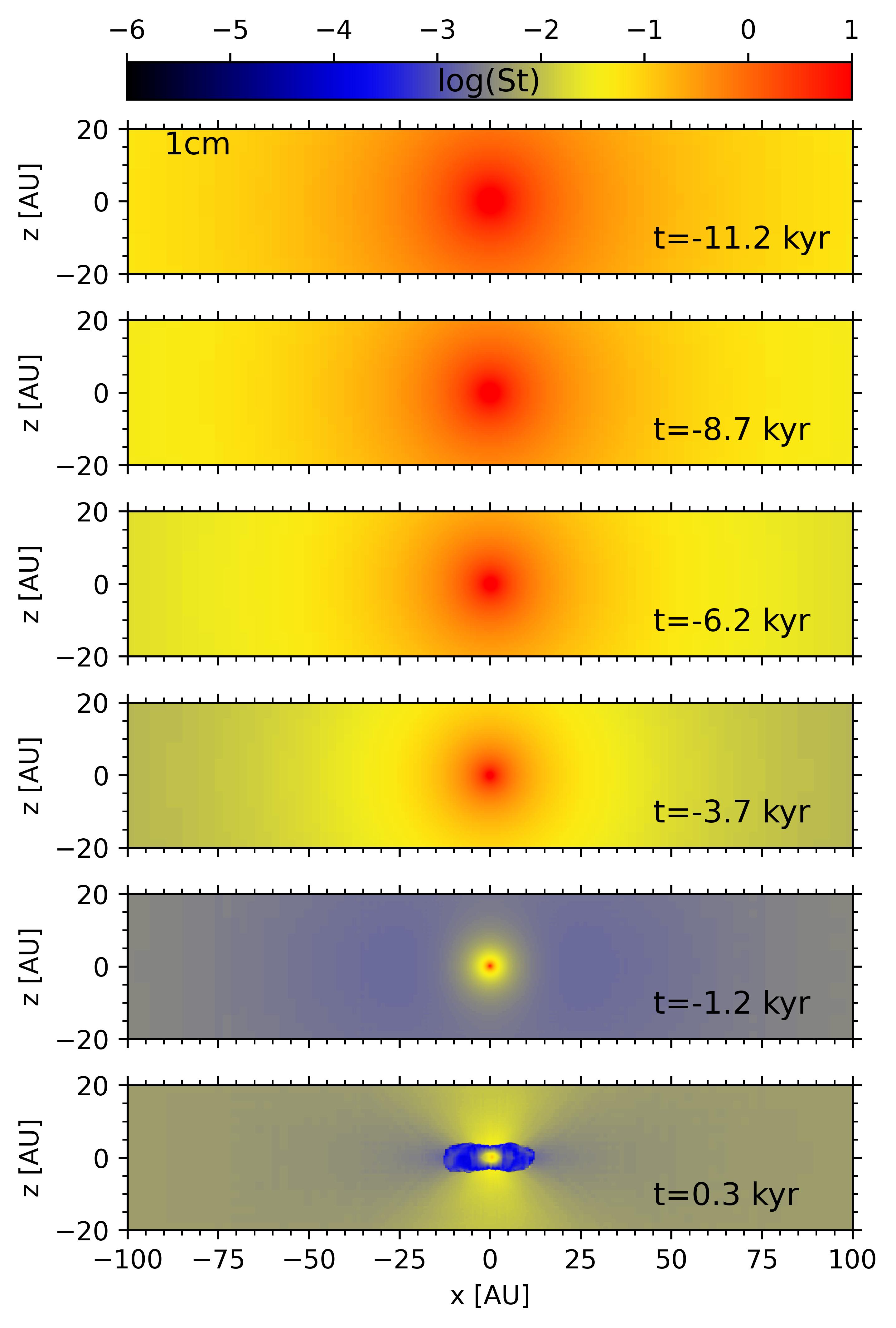}
    \caption{Time evolution of $\mathrm{St}$ in the $x-z$ slice. Left- and right-hand side group of panels belong to the models with $a_{\rm max}^{\rm cap}=10$~$\mu$m and $a_{\rm max}^{\rm cap}=1.0$~cm, respectively. Time is counted from the disk formation instance }
   \label{fig:6plot_xz_dens_dust_10}
\end{figure*}

Figure~\ref{fig:6plot_xz_dens_dust_10} presents the Stokes number in the vertical $x-z$ cut through the $y=0$ plane. The time period immediately preceding disk formation is shown.   Clearly, the Stokes number is larger in the inner regions of the collapsing cloud, exceeding unity in its center. The bottom row shows a time instance right after the FHSC formation. The Stokes number decreases in this stage owing to a sharp increase in the gas volume density and temperature. We note that the Stokes number definition for the envelope may not be fully self-consistent, see the main text.

\section{Dust growth rates in the envelope}
\label{App:four}
Equation~\ref{turb_vel} defines the turbulence-induced collision velocities in disk conditions, which may deviate from those in the surrounding protostellar envelope. The original formulation from \citet{2007OrmelCuzzi} requires specifying the velocity~($v_{\rm L}$) and spatial scales~($L$) of the largest turbulent eddy. In the $\alpha$-prescription for disk turbulence, these scales are proportional to the local sound speed $c_{\rm s}$ and the gas vertical scale height $H_{\rm gas}=c_{\rm s}/\Omega_{K}$ (e.g., \citealt{2024ApJ...966...90S}):
 \begin{eqnarray}
 v_{\rm L}&=&\sqrt{\alpha}c_{\rm s},\\
 L&=&\sqrt{\alpha}H_{\rm gas}.   
 \end{eqnarray}

The turbulence in the infalling envelope, if developed, may exhibit distinct properties compared to that in the disk. To quantify the influence of envelope turbulence on dust evolution, we compute the dust growth timescales employing the turbulence-induced velocity recipe from \citet{2007OrmelCuzzi}. We consider four parameter sets of $v_{\rm L}$ and $L$ values, and assume the same physical structure of the collapsing cloud at 3.7~kyr as in Figure~\ref{fig:growth-source}.

 \begin{eqnarray}
     v_{\rm L}&=&0.03c_{\rm s}, \quad L = 0.5\,{\rm au};\\
     v_{\rm L}&=&0.5c_{\rm s}, \quad L = 100\,{\rm au};\\
     v_{\rm L}&=&0.5c_{\rm s}, \quad L = 1000\,{\rm au};\\
     v_{\rm L}&=&1.0c_{\rm s}, \quad L = 1000\,{\rm au};\\
 \end{eqnarray}

\begin{figure}
    \centering
    \includegraphics[width=1.0\columnwidth]{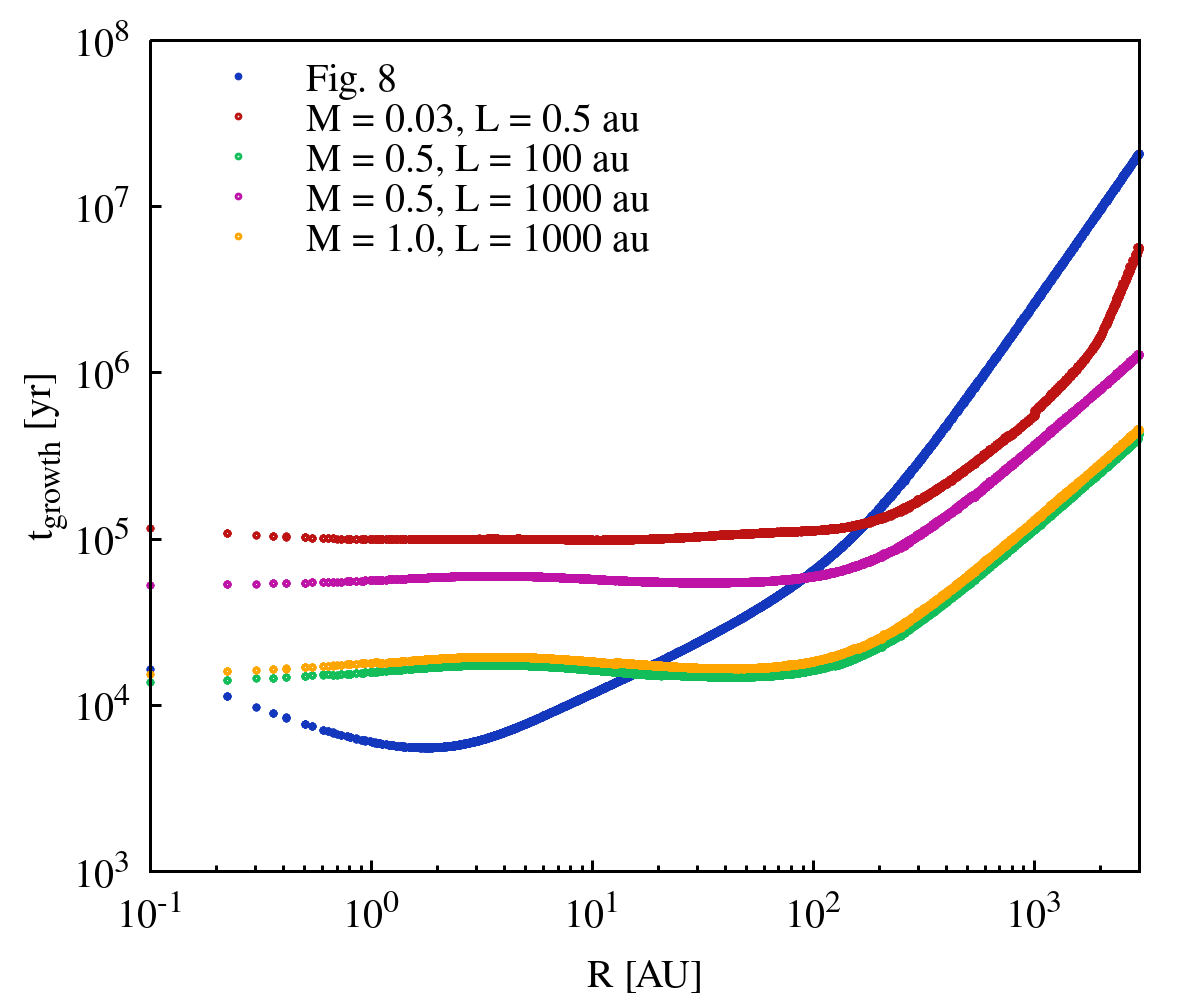}
    \caption{Dust growth timescales in the collapsing cloud for different turbulence properties in the envelope, parametrized by the Mach number~$M = v_{\rm L}/c_{\rm s}$ and the spatial scale~$L$ of the largest turbulent eddy.}
   \label{fig:DGRenv}
\end{figure}

The first case corresponds to the turbulence properties that are characteristic of the outer disk when it forms, while the next three cases explore the parameter space relevant for the envelopes \citep{2005MNRAS.357..687W, 2023ApJ...944..222S}. Figure~\ref{fig:DGRenv} demonstrates the comparison between the dust growth timescale calculated based on a simplified recipe for disk conditions (Equation~\ref{turb_vel}; blue line) and those for a stronger turbulence in the envelope. One can see that at the outer radii $R>10-100$~au, the stronger turbulence in the envelope may lead to even faster growth rates than those shown in Figure~\ref{fig:growth-source}. Thus, our key conclusion that dust growth starts in the contracting cloud in the evolution stage that precedes disk formation remains generally unchanged. We should note, however, that the value of $t_{\rm growth}$ is only an estimate as the dust growth rates depend on the grain size itself. More detailed modeling of the dust size evolution requires proper consideration of the turbulence changes at the envelope-disk interface, which deserves a separate study.

\end{appendix}


\end{document}